\documentclass[twocolumn,showpacs,preprintnumbers,amsmath,amssymb,fleqn,showkeys]{revtex4}

\usepackage{graphicx}
\usepackage{dcolumn}
\usepackage{bm}

\usepackage{color}
\usepackage{epsfig}

\newcommand{\gtapprox}{\raisebox{-0.5ex}{$\,\stackrel{>}{\scriptstyle\sim}\,$}}
\newcommand{\ltapprox}{\raisebox{-0.5ex}{$\,\stackrel{<}{\scriptstyle\sim}\,$}}

\begin{document}



\title{Classes of confining gauge field configurations}

\author{Marc Wagner}

\affiliation{Institute~for~Theoretical~Physics~III, University~of~Erlangen-N{\"u}rnberg, Staudtstra{\ss}e~7, 91058~Erlangen, Germany}

\date{\today}

\begin{abstract}
We present a numerical method to compute path integrals in effective SU(2) Yang-Mills theories. The basic idea is to approximate the Yang-Mills path integral by summing over all gauge field configurations, which can be represented as a linear superposition of a small number of localized building blocks. With a suitable choice of building blocks many essential features of SU(2) Yang-Mills theory can be reproduced, particularly confinement. The analysis of our results leads to the conclusion that topological charge as well as extended structures are essential elements of confining gauge field configurations.
\end{abstract}



\maketitle


\section{Introduction}

A common approach to get some insight regarding the mechanism of confinement is to approximate the Yang-Mills path integral by restricting the integration to certain subsets of gauge field configurations. By considering such effective theories one might find out, which gauge field configurations are responsible for confinement and which are not. Examples are ensembles of singular gauge instantons (c.f.\ e.g.\ \cite{Schafer:1996wv}), ensembles of calorons (c.f.\ e.g.\ \cite{Gross:1980br,Gerhold:2006sk}), ensembles of regular gauge instantons and ensembles of merons \cite{Lenz:2003jp,Negele:2004hs} or the removal of center vortices in lattice calculations (c.f.\ e.g.\ \cite{DelDebbio:1998uu}). Some of these approaches have analytical aspects but most of them extensively resort to numerical methods.

In this work we generalize the ideas and techniques presented in \cite{Lenz:2003jp,Negele:2004hs}. We study different classes of gauge field configurations, especially their importance with regard to confinement, by applying a numerical method called pseudoparticle approach \cite{Wagner:2005vs,Wagner:2006,Wagner:2006du}. We demonstrate that with a suitable choice of building blocks many essential features of SU(2) Yang-Mills theory can be reproduced, particularly confinement.

This paper is organized as follows.
In section~\ref{SEC_001} we explain the basic principle of the pseudoparticle approach. We also discuss numerical issues and compare the method to lattice gauge theory.
In section~\ref{SEC_003} we show how to calculate certain observables in the pseudoparticle approach: the static quark antiquark potential at zero and at finite temperature, the topological susceptibility and the critical temperature of the confinement deconfinement phase transition. We present numerical results, which are in qualitative agreement with lattice results.
In section~\ref{SEC_007} we apply the pseudoparticle approach with different types of building blocks, to study different classes of gauge field configurations and their effect on confinement. By doing that we try to determine properties of confining gauge field configurations. Our findings indicate that topological charge as well as extended structures are necessary to obtain confinement.
In section~\ref{SEC_011} we summarize our results and give a brief outlook regarding future research.


\section{\label{SEC_001}The pseudoparticle approach}


\subsection{SU(2) Yang-Mills theory}

In this work we consider Euclidean SU(2) Yang-Mills theory. The action is given by
\begin{eqnarray}
S \ \ = \ \ \int d^4x \, s
\end{eqnarray}
with action density
\begin{eqnarray}
s \ \ = \ \ \frac{1}{4 g^2} F_{\mu \nu}^a F_{\mu \nu}^a
\end{eqnarray}
($g$ is the dimensionless coupling constant) and field strength
\begin{eqnarray}
F_{\mu \nu}^a \ \ = \ \ \partial_\mu A_\nu^a - \partial_\nu A_\mu^a + \epsilon^{a b c} A_\mu^b A_\nu^c
\end{eqnarray}
($\epsilon_{1 2 3} = 1$). The topological charge is defined by
\begin{eqnarray}
Q \ \ = \ \ \int d^4x \, q
\end{eqnarray}
with topological charge density
\begin{eqnarray}
q \ \ = \ \ \frac{1}{32 \pi^2} F_{\mu \nu}^a \tilde{F}_{\mu \nu}^a
\end{eqnarray}
and dual field strength
\begin{eqnarray}
\tilde{F}_{\mu \nu}^a \ \ = \ \ \frac{1}{2} \epsilon_{\mu \nu \alpha \beta} F_{\alpha \beta}^a
\end{eqnarray}
($\epsilon_{0 1 2 3} = -1$).


\subsection{The basic principle of the pseudoparticle approach}

The basic idea of the pseudoparticle approach is to restrict the Yang-Mills path integral to those gauge field configurations, which can be represented as a linear superposition of a small number of building blocks, typically around $400$. The building blocks are gauge field configurations, which are localized in Euclidean spacetime. We refer to these building blocks as pseudoparticles, where the term pseudoparticle refers to the fact that the corresponding gauge field configurations are also localized in time in contrast to ordinary particles, which are only localized in space. In our context a pseudoparticle is not necessarily a solution of the classical Yang-Mills equations of motion.

Every pseudoparticle or building block has certain parameters, a position, an amplitude and a color orientation, which uniquely define the resulting gauge field configuration. It is important to stress that in general a sum of such pseudoparticles is not even close to a classical solution, as it is the case e.g.\ in instanton gas models (c.f.\ e.g.\ \cite{Schafer:1996wv}). The pseudoparticle approach is supposed to describe full quantum physics and not only certain semiclassical corrections.

The integration over all gauge field configurations in the path integral is replaced by an integration over the amplitudes and color orientations of the pseudoparticles. When considering a spacetime region containing a finite number of pseudoparticles, this is an ordinary multidimensional integral, which can be computed by Monte-Carlo methods.

The starting point of this work has been \cite{Lenz:2003jp,Negele:2004hs}. However, there are two important generalizations:
\begin{itemize}
\item We do not restrict our approach to regular gauge instantons and merons. A pseudoparticle can be any localized gauge field configuration. For example we also employ pseudoparticles with a limited range of interaction or without topological charge.

\item In addition to a color orientation matrix we assign to each pseudoparticle a variable amplitude. In this way pseudoparticles are also able to model small quantum fluctuations.
\end{itemize}

Note that the pseudoparticle approach is a general technique, which is in no way restricted to SU(2) Yang-Mills theory. With minor modifications it can also be applied to other quantum field theories, e.g.\ quantum mechanics (c.f.\ \cite{Wagner:2006}).


\subsection{The standard choice of pseudoparticles: ``instantons'', ``antiinstantons'' and akyrons}

For the major part of this work we consider the following pseudoparticles:
\begin{eqnarray}
\nonumber & & \hspace{-0.44cm} A_\mu^a(x) \ \ = \ \ \mathcal{A}(i) \mathcal{C}^{a b}(i) a_{\mu,\textrm{instanton}}^b(x-z(i)) \quad , \\
\label{EQN_001} & & a_{\mu,\textrm{instanton}}^b(x) \ \ = \ \ \eta_{\mu \nu}^b \frac{x_\nu}{x^2 + \lambda^2} \\
\nonumber & & \hspace{-0.44cm} A_\mu^a(x) \ \ = \ \ \mathcal{A}(i) \mathcal{C}^{a b}(i) a_{\mu,\textrm{antiinstanton}}^b(x-z(i)) \quad , \\
\label{EQN_002} & & a_{\mu,\textrm{antiinstanton}}^b(x) \ \ = \ \ \bar{\eta}_{\mu \nu}^b \frac{x_\nu}{x^2 + \lambda^2} \\
\nonumber & & \hspace{-0.44cm} A_\mu^a(x) \ \ = \ \ \mathcal{A}(i) \mathcal{C}^{a b}(i) a_{\mu,\textrm{akyron}}^b(x-z(i)) \quad , \\
\label{EQN_003} & & a_{\mu,\textrm{akyron}}^b(x) \ \ = \ \ \delta^{b 1} \frac{x_\mu}{x^2 + \lambda^2} ,
\end{eqnarray}
where $\eta_{\mu \nu}^b = \epsilon_{b \mu \nu} + \delta_{b \mu} \delta_{0 \nu} - \delta_{b \nu} \delta_{0 \mu}$ and \\ $\bar{\eta}_{\mu \nu}^b = \epsilon_{b \mu \nu} - \delta_{b \mu} \delta_{0 \nu} + \delta_{b \nu} \delta_{0 \mu}$. Each pseudoparticle has an index $i$, an amplitude $\mathcal{A}(i) \in \mathbb{R}$, a color orientation matrix $\mathcal{C}^{ab}(i) \in \textrm{SO(3)}$, a position $z(i) \in \mathbb{R}^4$ and a size $\lambda \in \mathbb{R}^+$. When considering a single pseudoparticle, a color orientation matrix is equivalent to a global gauge transformation. Since such a global gauge transformation can be specified by an element of SU(2), for which $S^3$ is a suitable parameter space, it can be expressed in terms of $(c_0(i),\ldots,c_3(i)) \in S^3$, i.e.\ $c_0(i)^2 + \mathbf{c}(i)^2 = 1$:
\begin{eqnarray}
\nonumber & & \hspace{-0.44cm} \mathcal{C}^{a b}(i) \ \ = \ \ \delta^{a b} \Big(c_0(i)^2 - \mathbf{c}(i)^2\Big) + 2 c_a(i) c_b(i) + \\
 & & \hspace{0.62cm} 2 \epsilon^{a b c} c_0(i) c_c(i)
\end{eqnarray}
(c.f.\ Appendix~\ref{SEC_012}). Note that Euclidean Lorentz transformations are equivalent to color rotations, when considering a single instanton or antiinstanton \cite{Jackiw:1976dw}, while a single akyron is form invariant under such Lorentz transformations. Therefore Euclidean Lorentz transformations have been included by considering arbitrary color orientations.

Setting $\mathcal{A}(i) = 2$ in (\ref{EQN_001}) yields an instanton in regular gauge \cite{Belavin:1975fg}. Although for $\mathcal{A}(i) \neq 2$ such pseudoparticles are not actually instantons, we nevertheless refer to them by that term. For $\mathcal{A}(i) = 2$ the action of an instanton is $S = 8 \pi^2 / g^2$, otherwise it is $S = \infty$. The topological charge is given by
\begin{eqnarray}
\label{EQN_004} Q \ \ = \ \ \frac{1}{4} \mathcal{A}(i)^2 \Big(3 - \mathcal{A}(i)\Big) .
\end{eqnarray}
With exception of a sign reversal in (\ref{EQN_004}) the same applies for antiinstantons (\ref{EQN_002}).

A single akyron \footnote{Ancient Greek: \textit{akyros} $=$ pure gauge (literally ``without effect'').} is a pure gauge, i.e.\ $S = 0$ and $Q = 0$. Note that for linear superpositions of akyrons $S \neq 0$ in general, due to the non-Abelian nature of SU(2). However, any such superposition has vanishing topological charge density (c.f.\ Appendix~\ref{SEC_013}).

A common and essential property of instantons, antiinstantons and akyrons is their long range nature. For large $|x|$ the corresponding gauge fields decrease like $1/|x|$. As a consequence these pseudoparticles have the ability to interact over large distances.


\subsubsection*{Why this particular choice of pseudoparticles?}

An important reason for considering pseudoparticles (\ref{EQN_001}) and (\ref{EQN_002}) is their similarity to regular gauge instantons and merons, which are known to exhibit confinement \cite{Lenz:2003jp,Negele:2004hs}.

We additionally include akyrons (\ref{EQN_003}) so that the gauge field has both a transverse part and a longitudinal part (superpositions of instantons and antiinstantons form transverse gauge fields, whereas superpositions of akyrons form longitudinal gauge fields; c.f.\ Appendix~\ref{SEC_13}). Furthermore, one can show that in the continuum limit, i.e.\ in the limit of infinitely many pseudoparticles, instantons (or antiinstantons) and akyrons form a basis of all gauge field configurations (c.f.\ Appendix~\ref{SEC_014}).

Finally, numerical studies with different types of pseudoparticles have shown that observables are not very sensitive to moderate changes in the definition of the pseudoparticles. For example replacing $1 / (x^2 + \lambda^2)$ by $(1/\lambda^2) \exp(-x^2 / 2 \lambda^2)$ in (\ref{EQN_001}) to (\ref{EQN_003}) hardly affects numerical results, if $\lambda$ is sufficiently large (c.f.\ section~\ref{SEC_009}). It seems that results in the pseudoparticle approach mainly depend on certain ``global pseudoparticle properties'', like their ability to interact over sufficiently large distances or whether they carry topological charge or not.


\subsection{\label{SEC_002}Pseudoparticle ensembles}

\begin{figure}[b]
\begin{center}
\includegraphics{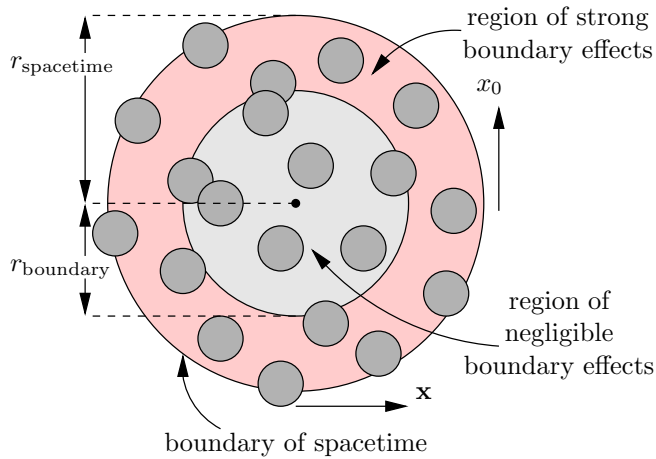}
\caption{\label{FIG_001}a 4-dimensional spacetime hypersphere of radius $r_\textrm{spacetime}$ with $N = 20$ pseudoparticles (pseudoparticles are symbolized by dark gray circles).}
\end{center}
\end{figure}

We put $N$ pseudoparticles with randomly and uniformly chosen positions inside a hyperspherical spacetime volume (c.f.\ Fig.~\ref{FIG_001}).
In the following we denote the radius of this spacetime hypersphere by $r_\textrm{spacetime}$ and its volume by $V_\textrm{spacetime}$. These quantities are related according to $V_\textrm{spacetime} = (\pi^2/2) r_\textrm{spacetime}^4$. The pseudoparticle density is given by $n = N / V_\textrm{spacetime}$. For most calculations we consider $N = 400$ pseudoparticles (\ref{EQN_001}) to (\ref{EQN_003}) with size $\lambda = 0.5$ and pseudoparticle density $n = 1.0$. This amounts to $r_\textrm{spacetime} = 3.0$.

The gauge field is a linear superposition of instantons, antiinstantons and akyrons:
\begin{eqnarray}
\nonumber & & \hspace{-0.44cm} A_\mu^a(x) \ \ = \ \ \sum_i \mathcal{A}(i) \mathcal{C}^{a b}(i) a_{\mu,\textrm{instanton}}^b(x-z(i)) + \\
\nonumber & & \hspace{0.62cm} \sum_j \mathcal{A}(j) \mathcal{C}^{a b}(j) a_{\mu,\textrm{antiinstanton}}^b(x-z(j)) + \\
\label{EQN_005} & & \hspace{0.62cm} \sum_k \mathcal{A}(k) \mathcal{C}^{a b}(k) a_{\mu,\textrm{akyron}}^b(x-z(k))
\end{eqnarray}
(the indices $i$, $j$ and $k$ assume values from different ranges). In accordance with the ratio of transverse and longitudinal gauge field components, which is $3 : 1$, we choose $N_\textrm{instanton} : N_\textrm{antiinstanton} : N_\textrm{akyron} = 3 : 3 : 2$ \\ ($N_\textrm{instanton}$, $N_\textrm{antiinstanton}$ and $N_\textrm{akyron}$ are the corresponding pseudoparticle numbers). Although in the limit of infinitely many pseudoparticles instantons by themselves form a basis of all transverse gauge fields, i.e.\ applying instantons and antiinstantons would yield an overcomplete basis, we consider an equal number of instantons and antiinstantons, when using a finite number of pseudoparticles. Our ensembles are then symmetric with respect to topological charge (c.f.\ (\ref{EQN_004})). 

We define the ensemble average of a quantity $\mathcal{O}$ by
\begin{eqnarray}
\nonumber & & \hspace{-0.44cm} \Big\langle \mathcal{O} \Big\rangle \ \ = \ \ \frac{1}{Z} \int \left(\prod_i d\mathcal{A}(i) \, d\mathcal{C}(i)\right) \mathcal{O}[A(\mathcal{A}(i),\mathcal{C}(i))] \\
\nonumber & & \hspace{0.62cm} e^{-S[A(\mathcal{A}(i),\mathcal{C}(i))]} \quad , \\
\label{EQN_006} & & Z \ \ = \ \ \int \left(\prod_i d\mathcal{A}(i) \, d\mathcal{C}(i)\right) e^{-S[A(\mathcal{A}(i),\mathcal{C}(i))]} \\
\label{EQN_007} & & \hspace{-0.44cm} S[A(\mathcal{A}(i),\mathcal{C}(i))] \ \ = \ \ \int_{V_\textrm{spacetime}} d^4x \, s(A(\mathcal{A}(i),\mathcal{C}(i)))
\end{eqnarray}
($d\mathcal{C}(i)$ is the invariant integration measure on $S^3$), i.e.\ the integration over all gauge field configurations in the path integral is replaced by an integration over the amplitudes and color orientations of the pseudoparticles. The quantity $\mathcal{O}$, the action $S$ and the action density $s$ can be expressed in terms of $\mathcal{A}(i)$ and $\mathcal{C}(i)$ via (\ref{EQN_005}). Only the action inside the spacetime hypersphere is considered for such a ``path integral''.

To eliminate the dependence of the ensemble average $\langle \mathcal{O} \rangle$ on the randomly chosen pseudoparticle positions $z(i)$, we independently calculate $\langle \mathcal{O} \rangle$ for many different sets of positions and take the average.

Instead of integrating over amplitudes and color orientations as in (\ref{EQN_006}) one could also integrate over positions or only over amplitudes. However, we have found that numerical results of such computations are very similar (c.f.\ \cite{Lenz:2003jp,Negele:2004hs,Wagner:2006}).

Note that our pseudoparticle ensembles are Lorentz invariant with respect to the center of the spacetime hypersphere as well as globally gauge invariant.


\subsubsection*{Numerical realization of pseudoparticle ensembles}

Ensemble averages $\langle \mathcal{O} \rangle$ in the pseudoparticle approach are given by multidimensional integrals (c.f.\ (\ref{EQN_006})), which can be computed by Monte-Carlo methods. We have applied the Metropolis algorithm (c.f.\ e.g.\ \cite{Rothe:2005nw}). In a single Metropolis step the pseudoparticles are updated one by one in fixed order. When updating a pseudoparticle, its amplitude and color orientation are randomly changed. These changes are always accepted if they reduce the action. Otherwise they may be rejected depending on the outcome of a stochastic experiment. The action inside the spacetime hypersphere (\ref{EQN_007}) is approximated by standard Monte-Carlo sampling.

\begin{figure}[h]
\begin{center}
\includegraphics{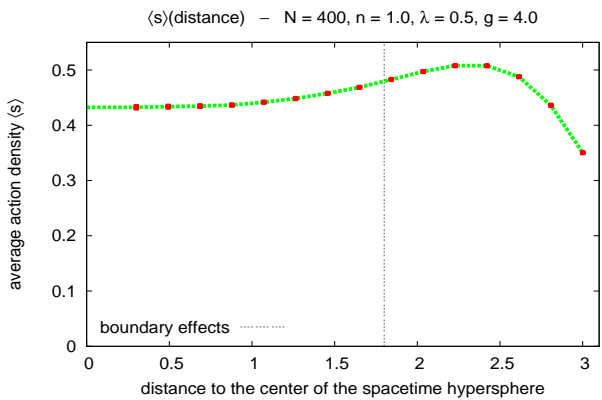}
\caption{\label{FIG_002}$N = 400$, $n = 1.0$, $\lambda = 0.5$, $g = 4.0$. $\langle s \rangle$ plotted against the distance to the center of the spacetime hypersphere.}
\end{center}
\end{figure}

To get rid of unwanted boundary effects, samples of physically meaningful quantities are always taken inside a hyperspherical spacetime region of radius \\ $r_\textrm{boundary} < r_\textrm{spacetime}$ (c.f.\ Fig.~\ref{FIG_001}). We determine \\ $r_\textrm{boundary}$ by considering plots of the average action density $\langle s \rangle$ against the distance to the center of the spacetime hypersphere: $r_\textrm{boundary}$ is assigned that distance, where $\langle s \rangle$ starts to deviate significantly from its constant value near the center of the spacetime hypersphere. In the example shown in Fig.~\ref{FIG_002} we have chosen $r_\textrm{boundary} = 1.8$.

Numerical calculations have shown that pseudoparticle results are pretty stable, when applying between $100$ and $800$ pseudoparticles (c.f.\ \cite{Wagner:2006}). Using significantly less than $100$ pseudoparticles is usually not possible, because to extract physically meaningful results one requires a sufficiently large spacetime region, where boundary effects are negligible.


\subsection{The pseudoparticle approach compared to lattice gauge theory}

In many aspects the pseudoparticle approach is similar to lattice gauge theory. Both approaches approximate path integrals by integrating over a finite number of degrees of freedom. In lattice gauge theory the integration is performed over a fixed number of link variables, while in the pseudoparticle approach there is a fixed number of pseudoparticles with amplitudes and color orientations as degrees of freedom.

In both cases the scale can be set by identifying the numerical value of any dimensionful quantity with its physical/experimental value. When using the string tension $\sigma$, any quantity $\mathcal{O}$ with dimension $(\textrm{length})^L$ is given in physical units by
\begin{eqnarray}
\label{EQN_008}\mathcal{O}_\textrm{physical} \ \ = \ \ \left(\frac{\sigma}{\sigma_\textrm{physical}}\right)^{L/2} \mathcal{O} ,
\end{eqnarray}
where $\mathcal{O}$ and $\sigma$ are the numerical dimensionless values of $\mathcal{O}$ and $\sigma$ at coupling constant $g$ and $\sigma_\textrm{physical}$ is the value of the string tension in physical units (throughout this work $\sigma_\textrm{physical} = 4.2 / \textrm{fm}^2$). A crucial property of any trustworthy numerical method is that dimensionless ratios of dimensionful quantities, e.g.\ $\chi^{1/4} / \sigma^{1/2}$ or $T_\textrm{critical} / \sigma^{1/2}$, do not depend on the coupling constant, i.e.\ that dimensionful quantities scale consistently. This has been observed both in lattice calculations and in the pseudoparticle approach (c.f.\ section~\ref{SEC_003}).

\begin{figure}[b]
\begin{center}
\includegraphics{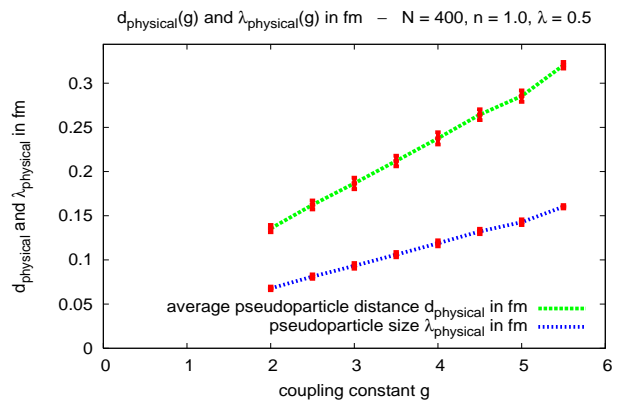}
\caption{\label{FIG_003}$N = 400$, $n = 1.0$, $\lambda = 0.5$. $\bar{d}_\textrm{physical}$ in $\textrm{fm}$ and $\quad \quad \quad$ $\lambda_\textrm{physical}$ in $\textrm{fm}$ plotted against $g$.}
\end{center}
\end{figure}

In lattice gauge theory the ultraviolet regulator is the lattice spacing $a$. Since different values for $g$ yield different values for $\sigma$, the lattice spacing in physical units $a_\textrm{physical}$ can be adjusted by choosing appropriate values for the coupling constant (replace $\mathcal{O}_\textrm{physical}$ by $a_\textrm{physical}$ and set $L=1$ and $a=1$ in (\ref{EQN_008})). In the pseudoparticle approach the minimum size of ultraviolet fluctuations is determined by the average pseudoparticle distance $\bar{d} = 1 / n^{1/4}$ and the pseudoparticle size $\lambda$, i.e.\ there are two ultraviolet regulators. A variation of the coupling constant $g$ has a similar effect on these regulators as it has in lattice calculations on the lattice spacing, that is $\bar{d}_\textrm{physical}$ and $\lambda_\textrm{physical}$ are increasing functions of $g$ (c.f.\ Fig.~\ref{FIG_003}; the scale has been set by the string tension, which we have obtained via generalized Creutz ratios (c.f.\ section~\ref{SEC_004})).

In contrast to lattice calculations we additionally have to specify the ratio $\lambda / \bar{d}$. For our standard choice of parameters, $\bar{d} = 1.0$ and $\lambda = 0.5$, the ultraviolet regulators are of the same order of magnitude. Numerical calculations have shown that physical quantities are pretty stable with respect to a variation of $\lambda / \bar{d}$ in the range $0.2 \ldots 1.1$ and that there is qualitative agreement with lattice results for suitably chosen pseudoparticles (c.f.\ section~\ref{SEC_008}).

When considering the limit of infinitely many pseudoparticles, any gauge field configuration can be represented by a suitable superposition of instantons and akyrons (c.f.\ section~\ref{SEC_014}), i.e.\ in this limit our pseudoparticle ensembles are gauge invariant. However, unlike in lattice gauge theory we only have approximate gauge invariance when working with a finite number of degrees of freedom. Whether this approximate gauge invariance is a problem can be tested by numerical calculations. A suitable way of doing that is to fix the gauge and to perform computations similar to those without gauge fixing. In the pseudoparticle approach a particularly convenient gauge is Lorentz gauge, i.e.\ $\partial_\mu A_\mu^a = 0$, which amounts to considering only instantons and antiinstantons but no akyrons. The qualitative agreement between these results and results without gauge fixing (c.f.\ section~\ref{SEC_010}) indicates that in the pseudoparticle approach the lack of exact gauge invariance is not a problem.

In contrast to lattice gauge theory we always work in the continuum, i.e.\ we do not discretize spacetime. This might be an advantage when calculating glueball masses from correlation functions, because due to exact rotational symmetry it is possible to project out states with a well defined angular momentum. Furthermore, the pseudoparticle approach  has a high potential when dealing with fermionic problems, because there is no fermion doubling in such a continuum formulation.

Another difference lies in the number of degrees of freedom, which is significantly smaller in the pseudoparticle approach than it is in typical lattice calculations ($1,600$ degrees of freedom, when using $400$ pseudoparticles, compared to e.g.\ $786,432$ degrees of freedom on a ``small'' $16^4$-lattice). Therefore, when pseudoparticle results are in agreement with results from lattice calculations, the pseudoparticles are chosen such that they represent essential degrees of freedom of SU(2) Yang-Mills theory. To put it another way, by applying different types of pseudoparticles one can determine the effect of certain classes of gauge field configurations on certain observables, and by doing that try to find out essential properties of confining gauge field configurations (c.f.\ section~\ref{SEC_007}). Lattice calculations, on the other hand, are much faster, i.e.\ they can cope with a significantly larger number of degrees of freedom in comparable computation time. Therefore, lattice calculations are certainly better suited for producing high quality numerical results.


\section{\label{SEC_003}Calculating observables}


\subsection{The static quark antiquark potential at zero temperature}

The common tool for studying the potential of a pair of infinitely heavy quarks at zero temperature are Wilson loops. A Wilson loop $W_z$ is defined by
\begin{eqnarray}
W_z[A] \ \ = \ \ \frac{1}{2} \textrm{Tr}\left(P\left\{\exp\left(i \oint dz_\mu \, A_\mu(z)\right)\right\}\right) ,
\end{eqnarray}
where $z$ is a closed spacetime curve and $P$ denotes path ordering. Rectangular Wilson loops with spatial extension $R$ and temporal extension $T$ are denoted by $W_{(R,T)}$.

It is well known that the potential of a static quark antiquark pair $V_{\textrm{q} \bar{\textrm{q}}}$ with separation $R$ can be related to ensemble averages of rectangular Wilson loops according to
\begin{eqnarray}
\label{EQN_009} V_{\textrm{q} \bar{\textrm{q}}}(R) \ \ = \ \ -\lim_{T \rightarrow \infty} \frac{1}{T} \ln \Big\langle W_{(R,T)} \Big\rangle
\end{eqnarray}
(c.f.\ e.g.\ \cite{Rothe:2005nw}).

In the following we assume that for large separations the static quark antiquark potential can be parameterized by
\begin{eqnarray}
\label{EQN_010} V_{\textrm{q} \bar{\textrm{q}}}(R) \ \ = \ \ V_0 - \frac{\alpha}{R} + \sigma R .
\end{eqnarray}
This ansatz is based on the bosonic string picture \cite{Luscher:1980fr,Luscher:1980ac} and on various numerical results from lattice calculations (c.f.\ e.g.\ \cite{Stack:1982wb,Bali:1994de,Luscher:2002qv}). There are three parameters:
\begin{itemize}
\item $V_0$ is a constant shift of the potential without physical relevance.

\item $\alpha$ is the coefficient in front of the attractive $1/R$-correction of the potential for large separations. The bosonic string picture predicts \\ $\alpha_\textrm{string} = \pi/12 \approx 0.26$ \cite{Luscher:1980fr,Luscher:1980ac}. In lattice calculations $\alpha_\textrm{lattice} = 0.22 \ldots 0.32$ has been observed \cite{Bali:1994de,Luscher:2002qv}. We refer to $\alpha$ as Coulomb coefficient.

\item The string tension $\sigma$ characterizes the force between a static quark and a static antiquark at large separations. Lattice calculations yield a positive value of the string tension, which implies the presence of confinement (c.f.\ e.g.\ \cite{Creutz:1980zw,Creutz:1980wj,Stack:1982wb,Bali:1994de}). Furthermore, $\sigma$ is a monotonically increasing function of the coupling constant $g$, i.e.\ when the scale is set by the string tension, the extension of the lattice in physical units can be adjusted by choosing appropriate values for $g$.
\end{itemize}


\subsubsection{\label{SEC_004}Calculating $\sigma$ and $\alpha$ via generalized Creutz ratios}

We determine the string tension $\sigma$ and the Coulomb coefficient $\alpha$ by generalizing the well known method of Creutz ratios \cite{Creutz:1980wj}.

Generalized Creutz ratios are based on guessing the functional dependence of ensemble averages of rectangular Wilson loops. A possible candidate is given by
\begin{eqnarray}
\nonumber & & \hspace{-0.44cm} -\ln \Big\langle W_{(R,T)} \Big\rangle \ \ = \\
\label{EQN_011} & & = \ \ V_0 \Big(R + T\Big) - \alpha \left(\frac{R}{T} + \frac{T}{R}\right) + \beta + \sigma R T
\end{eqnarray}
\cite{Bali:1994de}. It is a simple and plausible choice, which is consistent with numerical results from lattice calculations. It also fulfills the following necessary requirements:
\begin{itemize}
\item $\langle W_{(R,T)} \rangle = \langle W_{(T,R)} \rangle$ (in Euclidean spacetime there is no difference between space and time).

\item $\lim_{T \rightarrow \infty} (-\ln \langle W_{(R,T)} \rangle) = V_{\textrm{q} \bar{\textrm{q}}}(R) T$ (c.f.\ (\ref{EQN_009}) and (\ref{EQN_010})).
\end{itemize}
Note that (\ref{EQN_011}) is only valid for $R,T \gtapprox \bar{d},\lambda$. Wilson loops, which are significantly smaller than the two ultraviolet regulators $\bar{d}$ and $\lambda$, are subject to strong cutoff effects.

When applying generalized Creutz ratios, the starting point is a set of ensemble averages of rectangular Wilson loops,
\begin{eqnarray}
\label{EQN_012} \left\{\Big\langle W_{(R_1,T_1)} \Big\rangle \ , \ \Big\langle W_{(R_2,T_2)} \Big\rangle \ , \ \ldots \ , \ \Big\langle W_{(R_n,T_n)} \Big\rangle\right\} ,
\end{eqnarray}
with at least a couple of different ratios $R_i/T_i$.

A generalized Creutz ratio $\Gamma_X$ is defined by
\begin{eqnarray}
\nonumber & & \hspace{-0.44cm} \Gamma_X(R_{i_1},T_{i_1},\ldots,R_{i_4},T_{i_4}) \ \ = \\
\nonumber & & = \ \ \Big\langle W_{(R_{i_1},T_{i_1})} \Big\rangle^{c_{1,X}} \Big\langle W_{(R_{i_2},T_{i_2})} \Big\rangle^{c_{2,X}} \\
 & & \hspace{0.62cm} \Big\langle W_{(R_{i_3},T_{i_3})} \Big\rangle^{c_{3,X}} \Big\langle W_{(R_{i_4},T_{i_4})} \Big\rangle^{c_{4,X}}
\end{eqnarray}
with weights $c_{1,X}=c_{1,X}(R_{i_1},T_{i_1},\ldots,R_{i_4},T_{i_4}), \ldots,$ \\ $c_{4,X}=c_{4,X}(R_{i_1},T_{i_1},\ldots,R_{i_4},T_{i_4})$, which will be specified below ($X$ denotes any of the constants $V_0$, $\alpha$, $\beta$ or $\sigma$). Inserting the Wilson loop ansatz (\ref{EQN_011}) yields
\begin{eqnarray}
\nonumber & & \hspace{-0.44cm} -\ln\Big(\Gamma_X(R_{i_1},T_{i_1},\ldots,R_{i_4},T_{i_4})\Big) \ \ = \\
\nonumber & & = \ \ V_0 \bigg(c_{1,X} \Big(R_{i_1} + T_{i_1}\Big) + \ldots + \\
\nonumber & & \hspace{1.25cm} c_{4,X} \Big(R_{i_4} + T_{i_4}\Big)\bigg) + \\
\nonumber & & \hspace{0.62cm} \alpha \bigg(c_{1,X} \left(-\frac{R_{i_1}}{T_{i_1}}-\frac{T_{i_1}}{R_{i_1}}\right) + \ldots + \\
\nonumber & & \hspace{1.25cm} c_{4,X} \left(-\frac{R_{i_4}}{T_{i_4}}-\frac{T_{i_4}}{R_{i_4}}\right)\bigg) + \\
\nonumber & & \hspace{0.62cm} \beta \bigg(c_{1,X} + \ldots + c_{4,X}\bigg) + \\
\label{EQN_013} & & \hspace{0.62cm} \sigma \bigg(c_{1,X} \Big(R_{i_1} T_{i_1}\Big) + \ldots + c_{4,X} \Big(R_{i_4} T_{i_4}\Big)\bigg) .
\end{eqnarray}
By solving a linear system the weights $c_{1,X},\ldots,c_{4,X}$ are chosen such that (\ref{EQN_013}) reduces to
\begin{eqnarray}
\label{EQN_014} -\ln\Big(\Gamma_X(R_{i_1},T_{i_1},\ldots,R_{i_4},T_{i_4})\Big) \ \ = \ \ X .
\end{eqnarray}

To determine $V_0$, $\alpha$, $\beta$ and $\sigma$, we consider all four element subsets of (\ref{EQN_012}). For every subset we calculate $V_0$, $\alpha$, $\beta$ and $\sigma$ via (\ref{EQN_014}). Although there will be certain fluctuations due to systematical and statistical errors, most of these estimates should be pretty similar.

We have found that estimates for $V_0$, $\alpha$, $\beta$ and $\sigma$ with large $c_{1,X}^2 + \ldots + c_{4,X}^2$ exhibit significantly stronger fluctuations than estimates with small $c_{1,X}^2 + \ldots + c_{4,X}^2$ (c.f.\ Fig.~\ref{FIG_004}). Therefore, we sort the estimates according to $c_{1,X}^2 + \ldots + c_{4,X}^2$ and only keep the ``smaller half''. The average of these estimates is the final result for $V_0$, $\alpha$, $\beta$ or $\sigma$.


\subsubsection*{Results: $N = 400$,  $n = 1.0$,  $\lambda = 0.5$}

We have computed ensemble averages of rectangular Wilson loops $\langle W_{(R,T)} \rangle$ for $R,T \in \{7a \, , \, 8a \, , \, \ldots \, , \, 12a\}$, $a = 0.21$, $r_\textrm{boundary} = 1.8$, and have determined all possible generalized Creutz ratios. There are around $6,000$ generalized Creutz ratios for each of the quantities $V_0$, $\alpha$, $\beta$ and $\sigma$.
\begin{figure}[b]
\begin{center}
\includegraphics{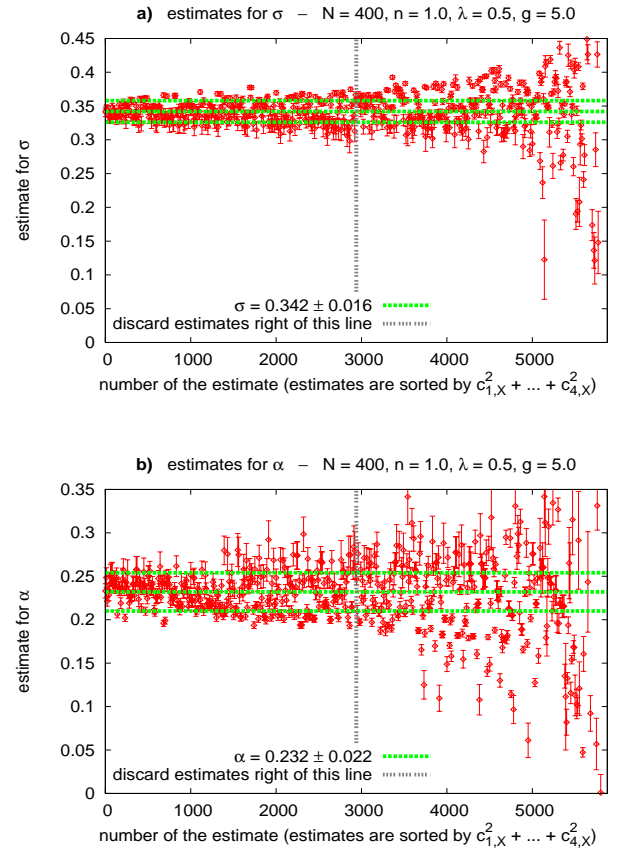}
\caption{\label{FIG_004}$N = 400$, $n = 1.0$, $\lambda = 0.5$, $g = 5.0$.
\textbf{a)}~Estimates for $\sigma$ sorted according to $c_{1,X}^2 + \ldots + c_{4,X}^2$.
\textbf{b)}~Estimates for $\alpha$ sorted according to $c_{1,X}^2 + \ldots + c_{4,X}^2$.
}
\end{center}
\end{figure}

In Fig.~\ref{FIG_004} the corresponding estimates for the string tension $\sigma$ and the Coulomb coefficient $\alpha$ are plotted for $g = 5.0$, sorted according to $c_{1,X}^2 + \ldots + c_{4,X}^2$ in increasing order from left to right (for the sake of clarity only every tenth estimate is shown). Both plots demonstrate that estimates on the left exhibit a smaller variance than estimates on the right. This indicates that estimates with small $c_{1,X}^2 + \ldots + c_{4,X}^2$ are more reliable than estimates with large $c_{1,X}^2 + \ldots + c_{4,X}^2$. To extract numerical values for $\sigma$ and $\alpha$, we have calculated the average and the standard deviation from the ``smaller half''. The results are $\sigma = 0.342 \pm 0.016$ and $\alpha = 0.232 \pm 0.022$.

By proceeding in the same way we have obtained numerical values for $\sigma$ and $\alpha$ for different coupling constants $g \in \{2.0 \, , \, 2.5 \, , \, \ldots \, , \, 5.5\}$.

\begin{figure}[h]
\begin{center}
\includegraphics{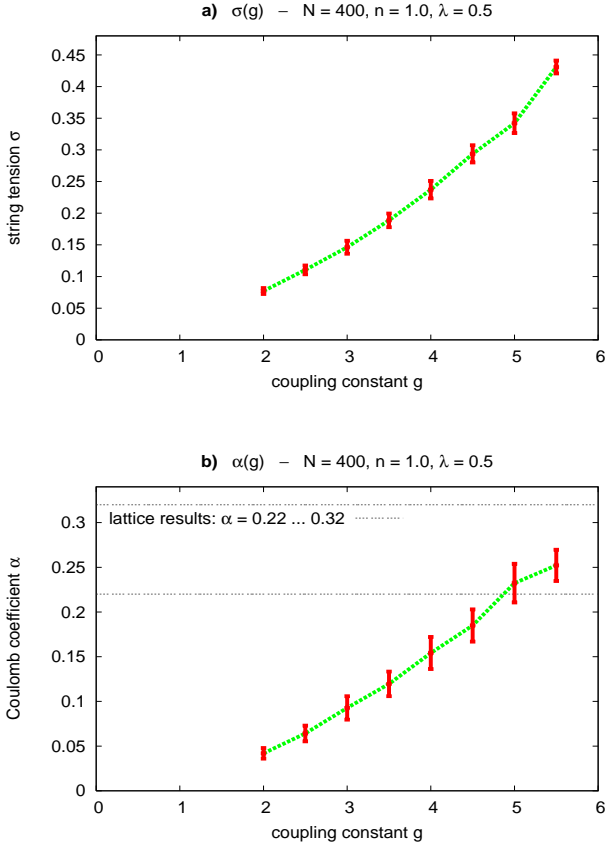}
\caption{\label{FIG_005}$N = 400$, $n = 1.0$, $\lambda = 0.5$.
\textbf{a)}~$\sigma$ plotted against $g$.
\textbf{b)}~$\alpha$ plotted against $g$.
}
\end{center}
\end{figure}

In Fig.~\ref{FIG_005}a the string tension $\sigma$ is plotted against the coupling constant $g$. It is positive and an increasing function of $g$. When the scale is set by the string tension, the size of the spacetime hypersphere in physical units can be adjusted by choosing appropriate values for the coupling constant. For $\sigma_\textrm{physical} = 4.2 / \textrm{fm}^2$ its diameter ranges from approximately $0.81 \, \textrm{fm}$ at $g = 2.0$ to $1.92 \, \textrm{fm}$ at \\ $g = 5.5$.

Fig.~\ref{FIG_005}b shows the Coulomb coefficient $\alpha$ as a function of the coupling constant $g$. It increases from $\alpha = 0.04$ at $g = 2.0$ to $\alpha = 0.25$ at $g = 5.5$. For large $g$ the value of $\alpha$ is in qualitative agreement with the prediction from the bosonic string picture, $\alpha_\textrm{string} = \pi/12 \approx 0.26$ \cite{Luscher:1980fr,Luscher:1980ac}, and with results from lattice calculations, \\ $\alpha_\textrm{lattice} = 0.22 \ldots 0.32$ \cite{Bali:1994de,Luscher:2002qv}.


\subsubsection{\label{SEC_005}Calculating the static quark antiquark potential}

The starting point to calculate the static quark antiquark potential is (\ref{EQN_009}), which we write as
\begin{eqnarray}
\label{EQN_015} V_{\textrm{q} \bar{\textrm{q}}}(R) T \ \ \approx \ \ -\ln \Big\langle W_{(R,T)} \Big\rangle .
\end{eqnarray}
We compute ensemble averages of rectangular Wilson loops $\langle W_{(R,T)} \rangle$ for fixed $R$ but different $T$ to obtain a curve $-\ln \langle W_{(R=\textrm{constant},T)} \rangle$.
\begin{figure}[b]
\begin{center}
\includegraphics{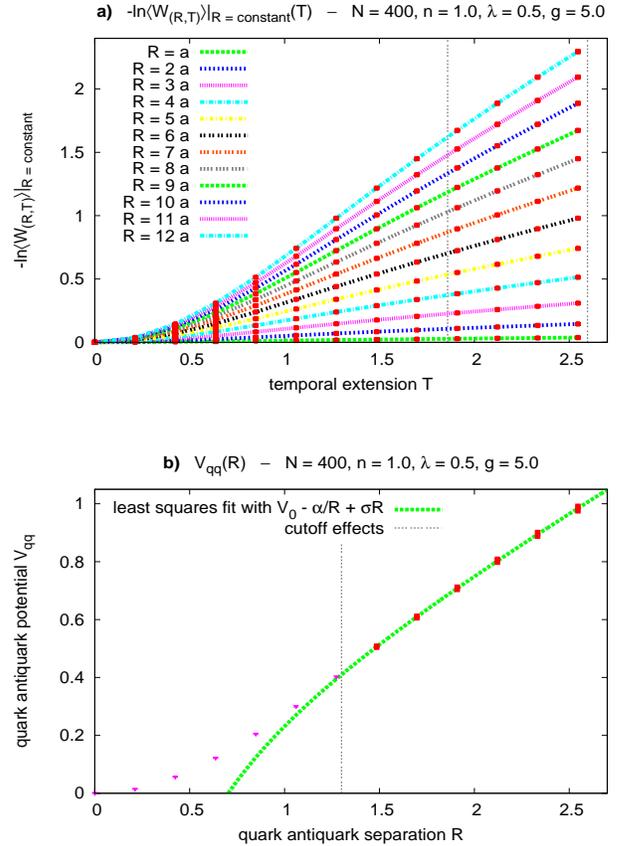}
\caption{\label{FIG_006}$N = 400$, $n = 1.0$, $\lambda = 0.5$, $g = 5.0$. \\
\textbf{a)}~$-\ln\langle W_{(R = \textrm{constant},T)} \rangle$ for different $R$ plotted against $T$ $\quad \quad$ ($a = 0.21$).
\textbf{b)}~$V_{\textrm{q} \bar{\textrm{q}}}$ plotted against the separation of the quarks.
}
\end{center}
\end{figure}
Examples are shown in Fig.~\ref{FIG_006}a. According to (\ref{EQN_015}) such a curve will exhibit a linear behavior for sufficiently large $T$. From the slope, which we obtain by fitting a straight line, we can read off $V_{\textrm{q} \bar{\textrm{q}}}(R)$. Iterating this procedure for a number of different $R$ yields an approximation of the quark antiquark potential.


\subsubsection*{Results: $N = 400$,  $n = 1.0$,  $\lambda = 0.5$,  $g = 5.0$}

We have computed ensemble averages of rectangular Wilson loops $\langle W_{(R,T)} \rangle$ for $R,T \in \{a \, , \, 2a \, , \, \ldots \, , \, 12a\}$, \\ $a = 0.21$, $r_\textrm{boundary} = 1.8$.

Fig.~\ref{FIG_006}a shows $-\ln \langle W_{(R=\textrm{constant},T)} \rangle$ as a function of $T$ for different $R \in \{a \, , \, 2a \, , \, \ldots \, ,\, 12a\}$. To determine the slope of these curves for large $T$, we have fitted straight lines to the data points at $T \in \{9a \, , \,  10a \, , \, 11a \, , \, 12a\}$, as indicated by the dashed straight lines.

The corresponding potential as a function of the quark antiquark separation is plotted in Fig.~\ref{FIG_006}b. For large separations it clearly exhibits a linear behavior. To obtain numerical values for the string tension $\sigma$ and the Coulomb coefficient $\alpha$, we have performed a least squares fit of the potential parameterization (\ref{EQN_010}) to the data points shown in Fig.~\ref{FIG_006}b. Only data points with $R \geq 1.3 > \bar{d},\lambda$ have been considered, because cutoff effects are expected to render the potential unphysical for small separations. Within statistical errors the results, $\sigma = 0.376 \pm 0.026$ and $\alpha = 0.283 \pm 0.093$, are in agreement with the results obtained by generalized Creutz ratios (c.f.\ section~\ref{SEC_004}). The fit is also shown in Fig.~\ref{FIG_006}b.

Note that in contrast to generalized Creutz ratios we have not made any assumption about the functional dependence of ensemble averages of Wilson loops to calculate the quark antiquark potential. Therefore, the agreement of the results for $\sigma$ and $\alpha$ with results obtained by generalized Creutz ratios shows again the consistency of the Wilson loop ansatz (\ref{EQN_011}) and Monte-Carlo data for $\langle W_{(R,T)} \rangle$.


\subsection{The topological susceptibility}

In order to produce quantitative results involving the string tension, we need other dimensionful quantities so that we can consider dimensionless ratios. One such quantity, which has been studied extensively on the lattice (c.f.\ e.g.\ \cite{DiVecchia:1981qi,Teper:1989ig,Christou:1995zn,DiGiacomo:1997wf}), is the topological susceptibility $\chi$. The topological susceptibility is closely related to the mass of the $\eta'$ meson \cite{Witten:1979vv}. It is defined by
\begin{eqnarray}
\chi \ \ = \ \ \lim_{V \rightarrow \infty} \frac{1}{V} \Big\langle Q_V^2 \Big\rangle ,
\end{eqnarray}
where $Q_V$ is the topological charge inside the spacetime volume $V$.

In our numerical calculations we approximate the limit $V \rightarrow \infty$ by a finite volume, the hyperspherical spacetime region with radius $r_\textrm{boundary}$ (c.f.\ section~\ref{SEC_002}). A number of computations has shown that $\chi$ is stable with respect to a a variation of that spacetime volume.


\subsubsection*{Results: $N = 400$, $n = 1.0$, $\lambda = 0.5$}

\begin{figure}[t]
\begin{center}
\includegraphics{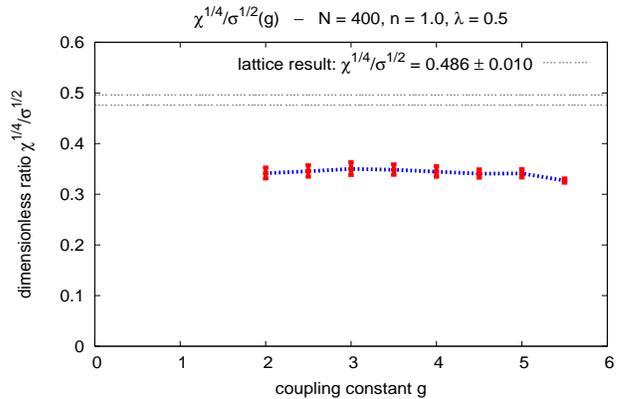}
\caption{\label{FIG_007}$N = 400$, $n = 1.0$, $\lambda = 0.5$. $\chi^{1/4}/\sigma^{1/2}$ plotted against $g$.}
\end{center}
\end{figure}

Fig.~\ref{FIG_007} shows the dimensionless ratio $\chi^{1/4}/\sigma^{1/2}$ as a function of the coupling constant $g$ ($r_\textrm{boundary} = 1.8$, $\sigma$ has been obtained by generalized Creutz ratios (c.f.\ Fig.~\ref{FIG_005}a)). As expected, $\chi^{1/4}/\sigma^{1/2}$ is nearly independent of $g$, i.e.\ the string tension and the topological susceptibility exhibit consistent scaling behaviors with respect to the coupling constant. This success strongly indicates that the pseudoparticle approach has the potential to reproduce correct Yang-Mills physics. The range of values, $\chi^{1/4}/\sigma^{1/2} = 0.33 \ldots 0.35$, is of the right order of magnitude compared to the lattice result \\ $(\chi^{1/4}/\sigma^{1/2})_\textrm{lattice} = 0.486 \pm 0.010$ \cite{Teper:1998kw}.


\subsection{The critical temperature of the confinement deconfinement phase transition and the finite temperature quark antiquark potential}


\subsubsection{\label{SEC_006}The pseudoparticle approach in finite temperature SU(2) Yang-Mills theory}

For finite temperature calculations we have to consider periodic gauge field configurations in $x_0$-direction. Therefore we need a method to generate periodic building blocks. In the case of singular gauge instantons this has already been done in form of calorons \cite{Harrington:1978ve,Gross:1980br}, which are periodic solutions to the classical Yang-Mills equations of motion. However, there is no straightforward generalization to arbitrary pseudoparticles. Therefore, we introduce a different method, which resorts to a blending technique from computer aided geometric design \cite{Wagner:2003}.

\begin{figure}[t]
\begin{center}
\includegraphics{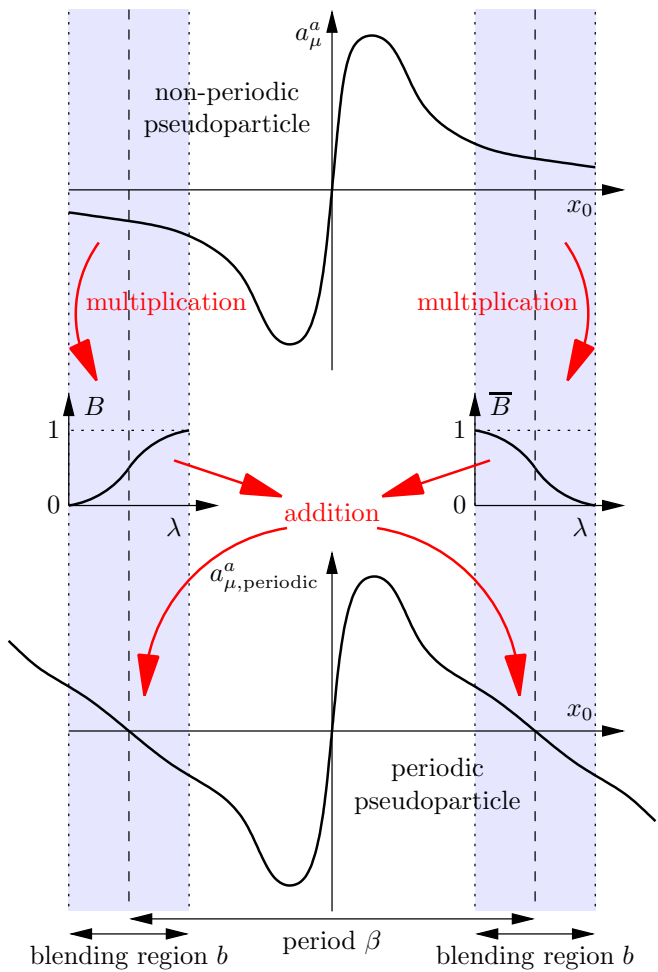}
\caption{\label{FIG_008}the basic principle of the blending method.}
\end{center}
\end{figure}

At first, we define blending functions $B$ and $\overline{B}$, which form smooth, $C^1$-continuous connections between $0$ at $\lambda = 0$ and $1$ at $\lambda = 1$ and vice versa:
\begin{eqnarray}
B(\lambda) \ \ = \ \ -2 \lambda^3 + 3 \lambda^2 \quad , \quad \overline{B}(\lambda) \ \ = \ \ 1 - B(\lambda) .
\end{eqnarray}
To make a non-periodic pseudoparticle $a_\mu^a$ with its center at the origin periodic in $x_0$-direction, we multiply ``both ends'' with the blending functions and add the results (c.f.\ Fig.~\ref{FIG_008}):
\begin{eqnarray}
\nonumber & & \hspace{-0.44cm} a_{\mu,\textrm{periodic}}^a(x) \ \ = \\
\nonumber & & = \ \ \left\{ \begin{array}{l} a_\mu^a(x_0,\mathbf{x}) \textrm{ if } -\frac{\beta-b}{2} \leq x_0 \leq \frac{\beta-b}{2} \\ \vspace{-0.2cm} \\ B(\lambda) a_\mu^a(x_0-\beta,\mathbf{x}) + \overline{B}(\lambda) a_\mu^a(x_0,\mathbf{x}) \\ \quad \quad \textrm{ if } \frac{\beta-b}{2} \leq x_0 \leq \frac{\beta+b}{2} \end{array} \right. , \\
\label{EQN_016} & & \lambda \ \ = \ \ \frac{x_0 - (\beta - b)/2}{b} .
\end{eqnarray}
To evaluate $a_{\mu,\textrm{periodic}}^a$ at $x_0 \notin [-(\beta-b)/2,(\beta+b)/2]$ one just has to combine (\ref{EQN_016}) and
\begin{eqnarray}
a_{\mu,\textrm{periodic}}^a(x_0,\mathbf{x}) = a_{\mu,\textrm{periodic}}^a(x_0+n \beta,\mathbf{x})
\end{eqnarray}
with a suitably chosen integer $n$. $b$ is the width of the blending region. Throughout this work we have chosen $b = 0.3 \times \beta$.


\subsubsection*{The spacetime region}

\begin{figure}[h!]
\begin{center}
\includegraphics{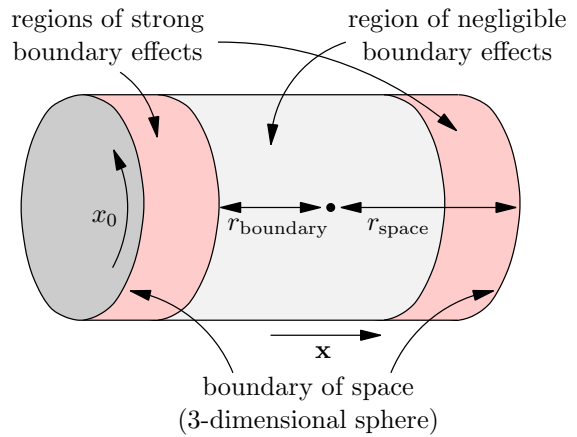}
\caption{\label{FIG_009}the spacetime region at finite temperature.}
\end{center}
\end{figure}

At zero temperature we consider a spacetime region, which is the interior of a $4$-dimensional hypersphere (c.f.\ Fig.~\ref{FIG_001}). At finite temperature that spacetime region is replaced by a spacetime with a periodic time direction of extension $\beta$ and a spatial part, which is the interior of an ordinary $3$-dimensional sphere of radius $r_\textrm{space}$ (c.f.\ Fig.~\ref{FIG_009}).

As in the zero temperature case we have to assure that samples of physically meaningful quantities are always taken inside a spacetime region, where boundary effects can be neglected. The spatial part of such a region is the interior of a sphere of radius $r_\textrm{boundary}$, whereas the time direction is not restricted (the light gray region in Fig.~\ref{FIG_009}).


\subsubsection*{The pseudoparticle density}

At finite temperature we use the same pseudoparticle density as for the corresponding zero temperature calculation (a zero temperature calculation with identical parameters is necessary to obtain a numerical value for the string tension, which is used to set the scale). This is a reasonable choice, because in the limit $\beta \rightarrow \infty$, \\ $r_\textrm{space} \rightarrow \infty$ and $r_\textrm{spacetime} \rightarrow \infty$ finite temperature ensembles and zero temperature ensembles become identical. Furthermore, there is a close analogy to lattice calculations. To set the scale, a zero temperature calculation with the same number of lattice sites in all four spacetime directions is carried out. For finite temperature calculations the number of lattice sites in the temporal direction is significantly reduced. However, the density of lattice sites is still the same, i.e.\ there is a smaller number of link variables in a smaller spacetime volume.

The pseudoparticle approach offers two possibilities to change the temperature:
\begin{itemize}
\item Varying $\beta$ while $g$ is kept constant changes the value of the temperature directly.

\item Varying $g$ while $\beta$ is kept constant changes the numerical value of the string tension $\sigma$. This leads to a different extension of the periodic time direction in physical units and, therefore, changes the value of the temperature.
\end{itemize}
Both methods yield consistent results (c.f.\ Fig.~\ref{FIG_010}c).


\subsubsection{The critical temperature of the confinement deconfinement phase transition}

Whereas at low temperatures static quarks cannot be separated from each other, at high temperatures a deconfining phase is expected. The temperature, at which the corresponding phase transition takes place, is called critical temperature and denoted by $T_\textrm{critical}$.

$T_\textrm{critical}$ can be determined from ensemble averages of Polyakov loops. A Polyakov loop is a Wilson loop around the periodic time $x_0$:
\begin{eqnarray}
\label{EQN_017} L_{\mathbf{z}}[A] \ \ = \ \ \frac{1}{2} \textrm{Tr}\left(P\left\{\exp\left(i \oint dz_0 \, A_0(z)\right)\right\}\right) .
\end{eqnarray}
Due to spatial translational invariance $\langle L_{\mathbf{z}} \rangle_\beta$ is $\mathbf{z}$-inde\-pendent. Therefore, from a numerical point of view it is convenient to consider spatial averages of Polyakov loops. We define
\begin{eqnarray}
\Big\langle L \Big\rangle_\beta \ \ = \ \ \left\langle\frac{1}{V} \int_V d^3z \, L_{\mathbf{z}}[A] \right\rangle_\beta .
\end{eqnarray}
$\langle L \rangle_\beta$ can be considered as an order parameter indicating whether there is confinement or not (c.f.\ e.g.\ \cite{Rothe:2005nw,McLerran:1981pb}):
\begin{eqnarray}
 & & \hspace{-0.44cm} \Big\langle L \Big\rangle_\beta \ \ = \ \ 0 \quad \leftrightarrow \quad \textrm{confinement} \\
 & & \hspace{-0.44cm} \Big\langle L \Big\rangle_\beta \ \ \neq \ \ 0 \quad \leftrightarrow \quad \textrm{deconfinement} .
\end{eqnarray}
This criterion is closely related to center symmetry, which is spontaneously broken in the deconfinement phase.

In the center symmetric phase of SU(2) Yang-Mills theory for every field configuration $A_\mu^{a,+}$ there is a gauge equivalent field configuration $A_\mu^{a,-}$ with \\ $L_\mathbf{z}[A^+] = -L_\mathbf{z}[A^-]$ ($A_\mu^{a,+}$ and $A_\mu^{a,-}$ are related by center symmetry, i.e.\ they are connected by a singular gauge transformation). Since both field configurations have the same action, i.e.\ $S[A^+] = S[A^-]$, $\langle L \rangle_\beta = 0$ follows immediately.

The spontaneous breakdown of center symmetry at $T = T_\textrm{critical}$ comes along with a splitting of the Hilbert space of states in two independent spaces: \\ $\mathcal{H} \rightarrow \mathcal{H}^+ \oplus \mathcal{H}^-$. The same applies for the set of field configurations considered in the path integral: \\ $\mathcal{A} \rightarrow \mathcal{A}^+ \oplus \mathcal{A}^-$. The ensemble average of the Polyakov loop depends on which Hilbert space was chosen during the spontaneous breakdown of center symmetry: $\langle L \rangle_{\beta,\mathcal{H}^+} = +\overline{l}$ and $\langle L \rangle_{\beta,\mathcal{H}^-} = -\overline{l}$.

In the broken phase two field configurations, which are related by center symmetry, cannot be connected continuously by a set of field configurations of finite action. This implies that during a Monte-Carlo simulation only field configurations either from $\mathcal{A}^+$ or from $\mathcal{A}^-$ are generated, assuming an infinite system and a local and continuous update mechanism. In numerical calculations these assumptions are only approximately fulfilled. Nevertheless, in lattice Monte-Carlo simulations (c.f.\ e.g.\ \cite{McLerran:1981pb,Gottlieb:1985ug}) it has been observed that there are long sequences of steps, where only field configurations corresponding to one of the two Hilbert spaces are generated.

Since our pseudoparticle ensembles are only approximately center symmetric, there is a smooth transition between the two phases. $\langle L \rangle_\beta \approx 0$ can be observed well below the critical temperature. For $\beta \approx \beta_{\textrm{critical}}$ the ensemble average of the Polyakov loop quickly rises to a non-zero value. For high temperatures $\langle L \rangle_\beta \approx 1$ (c.f.\ Fig.~\ref{FIG_010}a). Therefore, we define the critical temperature $T_\textrm{critical}$ (or equivalently its inverse $\beta_\textrm{critical}$) to be that temperature, where the ensemble average of the Polyakov loop crosses a certain value $\xi$, i.e.\
\begin{eqnarray}
\Big\langle L \Big\rangle_{\beta_\textrm{critical}} \ \ = \ \ \xi .
\end{eqnarray}


\subsubsection*{$\langle L \rangle_\beta$ in the pseudoparticle approach versus $\langle L \rangle_\beta$ in lattice calculations}

The main difference between our results and lattice results is that even in the deconfinement phase we never have observed $\langle L \rangle_\beta < 0$. We conclude that in the pseudoparticle approach the low action field configurations corresponding to $\mathcal{H}^-$ are underrepresented, i.e.\ there are more low action field configurations corresponding to $\mathcal{H}^+$ than to $\mathcal{H}^-$. This bias always forces the system to chose the Hilbert space $\mathcal{H}_+$, when center symmetry is spontaneously broken.

The bias can be explained by the following qualitative argument: for field configurations close to zero $L_\mathbf{z} \approx 1$ (c.f.\ (\ref{EQN_017})). Therefore, we expect these field configurations to be elements of $\mathcal{A}^+$. Furthermore, all of these field configurations are low action field configurations, which contribute significantly to the path integral. The set of these field configurations is denoted by $\mathcal{A}_{A \approx 0}$. Loosely speaking, all other low action field configurations are large enough so that $L_\mathbf{z}$ can pick up exponents, which are significantly different from zero. They are either elements of $\mathcal{A}^+$ or of $\mathcal{A}^-$. The set of these field configurations is denoted by $\mathcal{A}_{A \gg 0}$. However, there is numerical evidence that in the pseudoparticle approach with around $400$ pseudoparticles most low action field configurations have small gauge fields, i.e.\ are elements of $\mathcal{A}_{A \approx 0}$. This is closely related to the fact that our ensembles are only approximately gauge invariant: a gauge transformed field configuration from $\mathcal{A}_{A \approx 0}$, which is a field configuration in $\mathcal{A}_{A \gg 0}$, can only be approximated, when using around $400$ pseudoparticles; such an approximation usually has a higher action. Therefore, there are proportionally more field configurations in $\mathcal{A}_{A \approx 0}$ than in $\mathcal{A}_{A \gg 0}$. This is manifested in form of a bias.


\subsubsection*{Results: $r_\textrm{space} = 3.00$, $n = 1.0$,  $\lambda = 0.5$}

\begin{figure}[t]
\begin{center}
\includegraphics{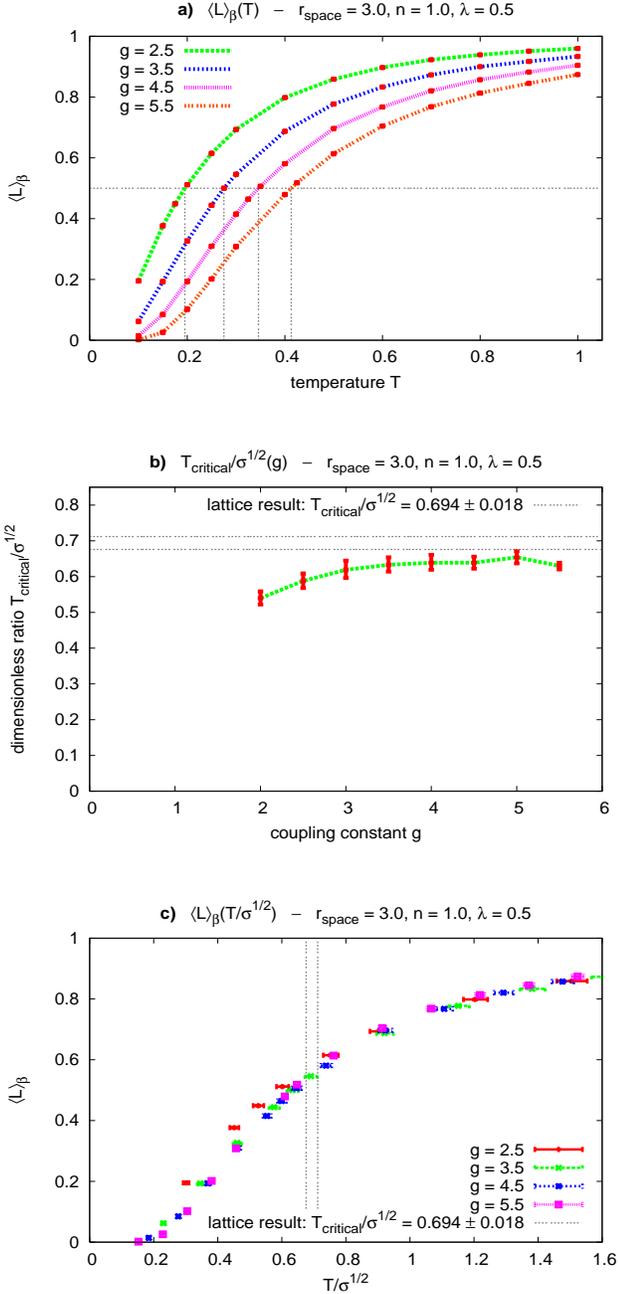}
\caption{\label{FIG_010}$r_\textrm{space} = 3.0$, $n = 1.0$, $\lambda = 0.5$.
\textbf{a)}~$\langle L \rangle_\beta$ for different $g$ plotted against $T$.
\textbf{b)}~$T_\textrm{critical}/\sigma^{1/2}$ plotted against $g$.
\textbf{c)}~$\langle L \rangle_\beta$ for different $g$ plotted against $T / \sigma^{1/2}$.
}
\end{center}
\end{figure}

In Fig.~\ref{FIG_010}a the ensemble average of the Polyakov loop $\langle L \rangle_\beta$ is plotted against the temperature $T$ for different coupling constants ($r_\textrm{space} = 3.00$, $r_\textrm{boundary} = 1.8$). The dashed straight lines correspond to a determination of $T_\textrm{critical}$ with $\xi = 0.5$ (other values for $\xi$ yield qualitatively identical results).

Fig.~\ref{FIG_010}b shows that the dimensionless ratio $T_\textrm{critical}/\sigma^{1/2}$ is nearly independent of the coupling constant $g$ ($\sigma$ has been obtained by generalized Creutz ratios (c.f.\ Fig.~\ref{FIG_005}a)), i.e.\ the string tension and the critical temperature scale consistently with respect to $g$. We would like to stress again that such a scaling behavior, although mandatory for any trustworthy numerical method, is far from obvious. The range of values, $T_\textrm{critical}/\sigma^{1/2} = 0.54 \ldots 0.65$, is of the right order of magnitude compared to the lattice result $(T_\textrm{critical}/\sigma^{1/2})_\textrm{lattice} = 0.694 \pm 0.018$ \cite{Teper:1998kw}.

As we have explained in section~\ref{SEC_006}, there are two methods to adjust the temperature, varying $\beta$ or varying $g$. Fig.~\ref{FIG_010}c shows  $\langle L \rangle_\beta$ as a function of $T / \sigma^{1/2}$ for different coupling constants $g$. The fact that all sample points scale to a single curve demonstrates that both methods yield consistent results.

\begin{figure}[h]
\begin{center}
\includegraphics{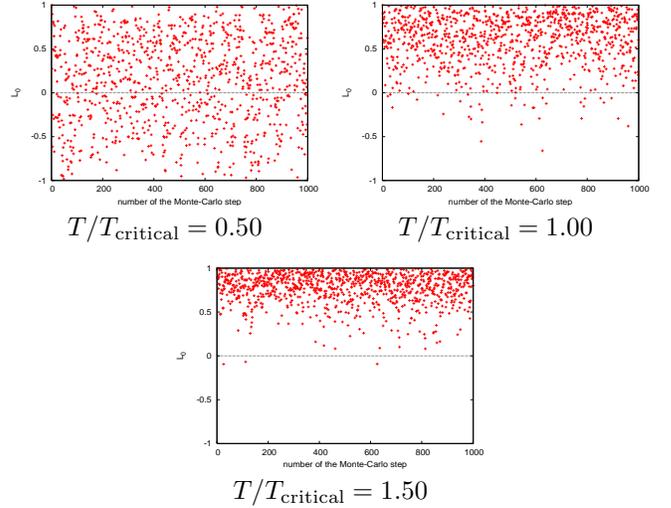}
\caption{\label{FIG_011}$r_\textrm{space} = 3.0$, $n = 1.0$, $\lambda = 0.5$, $g = 5.0$. $L_{\vec{0}}$ for different ratios $T / T_\textrm{critical}$ plotted against the number of the Monte-Carlo step.}
\end{center}
\end{figure}

Fig.~\ref{FIG_011} shows the evolution of $L_{\vec{0}}$ during a single Monte-Carlo simulation ($1,000$ Monte-Carlo steps) for $g = 5.0$ and different $T/T_\textrm{critical}$ ($T_\textrm{critical} = 0.38$). These plots demonstrate that in the confinement phase \\ ($T < T_\textrm{critical}$) $L_\mathbf{z}$ assumes approximately an equal number of positive and negative values, whereas in the deconfinement phase ($T > T_\textrm{critical}$) the values of $L_\mathbf{z}$ are mainly positive.


\subsubsection{The static quark antiquark potential at finite temperature}

It is well known that the static quark antiquark potential at finite temperature can be obtained from Polyakov loop correlation functions:
\begin{eqnarray}
\label{EQN_018} V_{\textrm{q} \bar{\textrm{q}}}(R,\beta) \ \ = \ \ -\frac{1}{\beta} \ln \Big\langle L_{\vec{0}} L_{\mathbf{z}}^\dagger \Big\rangle_\beta \quad , \quad |\mathbf{z}| = R
\end{eqnarray}
(c.f.\ e.g.\ \cite{McLerran:1981pb,Rothe:2005nw}).

However, center symmetry is only approximately realized in the pseudoparticle approach, i.e.\ even in the confinement phase $\langle L \rangle_\beta \neq 0$. Consequently, any finite temperature potential calculated according to (\ref{EQN_018}) approaches a constant value for large quark antiquark separations:
\begin{eqnarray}
\lim_{R \rightarrow \infty} V_{\textrm{q} \bar{\textrm{q}}}(R,\beta) \ \ = \ \ -\frac{2}{\beta} \ln \Big|\Big\langle L \Big\rangle_\beta\Big| .
\end{eqnarray}
Therefore, at finite temperature one can expect to observe confining potentials only up to intermediate separations.


\subsubsection*{Results: $r_\textrm{space} = 3.00$, $n = 1.0$,  $\lambda = 0.5$, $g = 5.0$}

\begin{figure}[h]
\begin{center}
\includegraphics{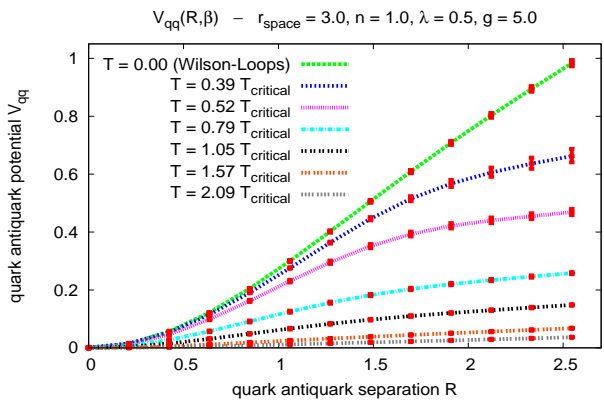}
\caption{\label{FIG_012}$r_\textrm{space} = 3.0$, $n = 1.0$, $\lambda = 0.5$, $g = 5.0$. Finite temperature potentials and the corresponding zero temperature potential plotted against the separation of the quarks.}
\end{center}
\end{figure}

Fig.~\ref{FIG_012} shows finite temperature quark antiquark potentials for $T/T_\textrm{critical} \in \{0.39 \, , \, 0.52 \, , \, \ldots \, , \, 2.09\}$ obtained from Polyakov loop correlation functions (c.f.\ (\ref{EQN_018}); \\ $r_\textrm{space} = 3.00$, $r_\textrm{boundary} = 1.27$) and the corresponding zero temperature potential obtained from ensemble averages of Wilson loops (c.f.\ section~\ref{SEC_005}). At high temperatures there is clearly no confinement. However, with the decrease of the ensemble average of the Polyakov loop towards lower temperatures finite temperature potentials gradually approach the zero temperature potential extracted from Wilson loops.

Because of $\langle L \rangle_\beta \neq 0$ the finite temperature potentials and the zero temperature potential differ at large separations even for temperatures significantly smaller than $T_\textrm{critical}$. There are arguments suggesting that these potentials should rather be calculated via
\begin{eqnarray}
\nonumber & & \hspace{-0.44cm} V_{\textrm{q} \bar{\textrm{q}}}(R,\beta) \ \ = \ \ -\frac{1}{\beta} \ln \Big\langle \Big(L_{\vec{0}} - \langle L \rangle_\beta\Big) \Big(L_{\mathbf{z}}^\dagger - \langle L \rangle_\beta\Big) \Big\rangle_\beta \quad , \\
 & & \hspace{0.62cm} |\mathbf{z}| = R
\end{eqnarray}
instead of (\ref{EQN_018}). Indeed, this yields finite temperature potentials, which are identical to the zero temperature potential within statistical errors, a result, which is in agreement with results from lattice calculations (c.f.\ e.g.\ \cite{Kaczmarek:1999mm}). We plan to discuss these topics in more detail in an upcoming paper.


\section{\label{SEC_007}Properties of confining gauge field configurations}

The goal of this section is to identify essential properties of gauge field configurations, which are responsible for confinement. To this end we compare different pseudoparticle ensembles.


\subsection{Pseudoparticles of different size and profile}


\subsubsection{\label{SEC_008}Pseudoparticles of different size}

In the following we explore how pseudoparticle results are affected by a variation of the pseudoparticle size $\lambda$, while the pseudoparticle density $n$ or equivalently the average pseudoparticle distance $\bar{d} = 1 / n^{1/4}$ is kept constant. In other words, we consider different ratios of the two ultraviolet regulators $\bar{d}$ and $\lambda$. Note that $\lambda$ strongly affects the shape of a pseudoparticle near its center, while it has essentially no effect on the long range behavior of a pseudoparticle, which is proportional to $1 / |x|$ (c.f.\ (\ref{EQN_001}) to (\ref{EQN_003})).

We consider ensembles with $N = 400$, $n = 1.0$, $g = 4.0$ and $\lambda \in \{0.20 \, , \, 0.35 \, , \, \ldots \, , \, 1.10\}$.

\begin{figure}[b]
\begin{center}
\includegraphics{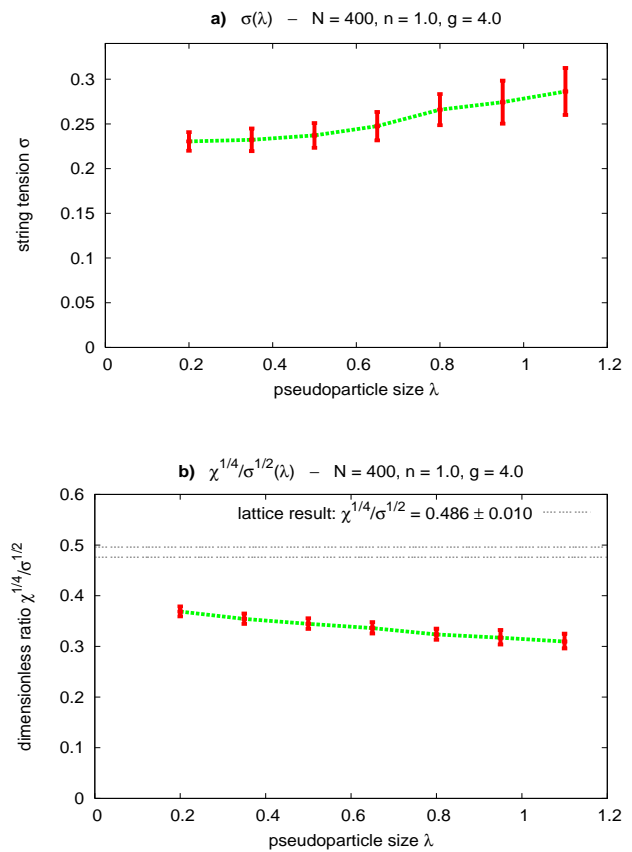}
\caption{\label{FIG_013}$N = 400$, $n = 1.0$, $g = 4.0$.
\textbf{a)}~$\sigma$ plotted against $\lambda$.
\textbf{b)}~$\chi^{1/4}/\sigma^{1/2}$ plotted against $\lambda$.
}
\end{center}
\end{figure}

We have obtained the string tension $\sigma$ via generalized Creutz ratios (c.f.\ section~\ref{SEC_004}). Fig.~\ref{FIG_013}a shows that $\sigma$ is clearly positive for all values of the pseudoparticle size $\lambda$. Moreover, there is only a weak $\lambda$-dependence, i.e.\ $\sigma = 0.23 \ldots 0.29$ for $\lambda = 0.20 \ldots 1.10$.
The implication is that confinement is connected to the $1/|x|$ long range behavior of the pseudoparticles, which is the same for all values of $\lambda$. The shape of the pseudoparticles near their center, which is $\lambda$-dependent, has essentially no effect on the string tension and, therefore, is of no relevance to confinement.

The topological susceptibility $\chi$ is also fairly independent of the pseudoparticle size $\lambda$ as is the dimensionless ratio $\chi^{1/4}/\sigma^{1/2}$ (c.f.\ Fig.~\ref{FIG_013}b). The range of values, $\chi^{1/4}/\sigma^{1/2} = 0.37 \ldots 0.31$ between $\lambda = 0.20 \ldots 1.10$, is in qualitative agreement with the lattice result \\ $(\chi^{1/4} / \sigma^{1/2})_\textrm{lattice} = 0.486 \pm 0.010$ \cite{Teper:1998kw}.


\subsubsection{\label{SEC_009}Pseudoparticles of Gaussian localized profile and different size}

To learn more about the interrelation between the long range behavior of pseudoparticles and confinement, we study pseudoparticles of Gaussian localized profile and different size. Such Gaussian localized pseudoparticles are obtained by replacing $1/ (x^2 + \lambda^2)$ appearing in the definitions of instantons, antiinstantons and akyrons ((\ref{EQN_001}) to (\ref{EQN_003})) by $(1/\lambda^2) e^{-x^2/(2 \lambda^2)}$. The main difference is the long range behavior of the resulting building blocks: in contrast to our standard choice of pseudoparticles, (\ref{EQN_001}) to (\ref{EQN_003}), Gaussian localized pseudoparticles have a limited range of interaction, which is proportional to their size $\lambda$.

We consider ensembles with $N = 400$, $n = 1.0$, $g = 4.0$ and $\lambda \in \{0.25 \, , \, 0.50 \, , \, \ldots \, , \, 1.50\}$.

\begin{figure}[h]
\begin{center}
\includegraphics{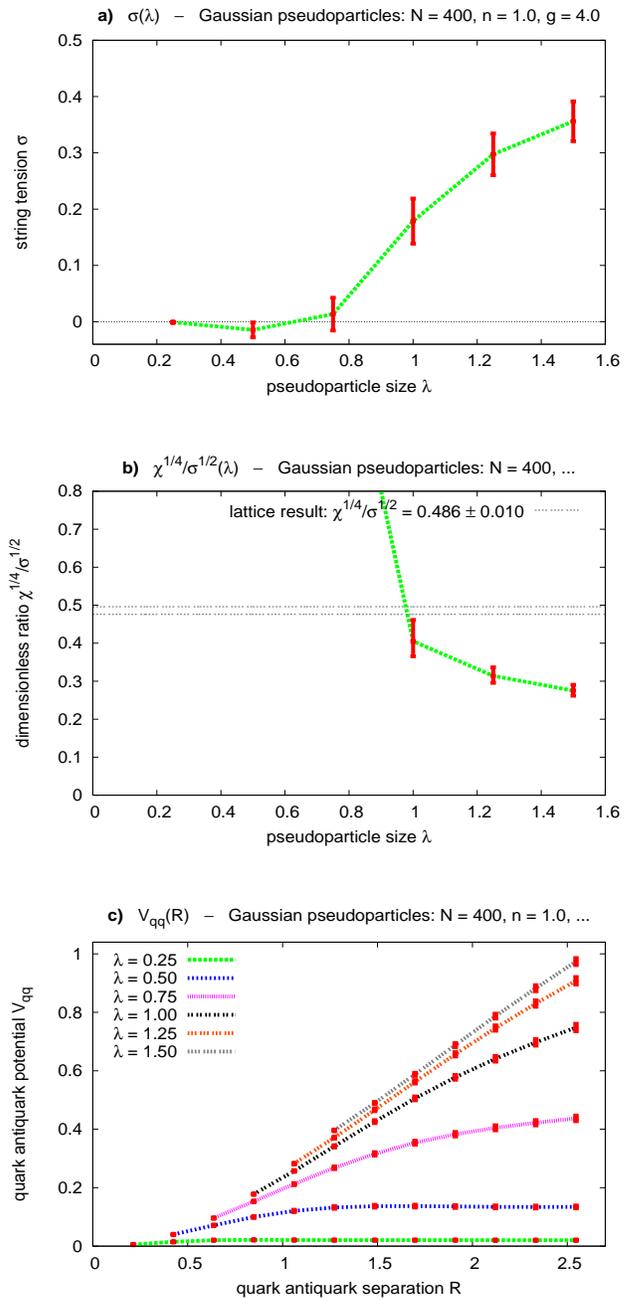}
\caption{\label{FIG_014}Gaussian localized pseudoparticles, $N = 400$, $n = 1.0$, $g = 4.0$.
\textbf{a)}~$\sigma$ plotted against $\lambda$.
\textbf{b)}~$\chi^{1/4} / \sigma^{1/2}$ plotted against $\lambda$.
\textbf{c)}~$V_{\textrm{q} \bar{\textrm{q}}}$ for different pseudoparticle sizes plotted against the separation of the quarks.
}
\end{center}
\end{figure}

We have obtained numerical values for the string tension $\sigma$ via generalized Creutz ratios (c.f.\ section~\ref{SEC_004}). The results are plotted in Fig.~\ref{FIG_014}a against the pseudoparticle size $\lambda$.
There is no confinement for $\lambda \ltapprox 0.75$ and confinement for $\lambda \gtapprox 1.00$. The onset of confinement takes place somewhere around $\lambda \approx 0.75$. This is precisely the width at which neighboring pseudoparticles start to overlap significantly (this can be seen by assigning each pseudoparticle an appropriate volume, e.g.\ the volume of a hypersphere of radius $\lambda$, $(\pi^2 / 2) \lambda^4 \approx 4.93 \times \lambda^4$, and comparing that volume with the maximum volume non-overlapping pseudoparticles can cover, i.e.\ $1/n = 1.0$). The implication is that pseudoparticle ensembles only exhibit confinement if their pseudoparticles cover sufficiently large spacetime regions so that there are significant overlaps between neighboring pseudoparticles. This is reminiscent of percolation. Note that percolation phenomena have been related to confinement in a variety of ways, e.g.\ percolation of center vortices \cite{Engelhardt:1999fd} or monopole loops \cite{Bornyakov:1991se} in lattice gauge theory or random percolation of bonds or sites on three dimensional lattices \cite{Gliozzi:2005ny}.

The $\lambda$-dependence of the topological susceptibility $\chi$ is much weaker than the $\lambda$-dependence of the string tension $\sigma$. Therefore, the dimensionless ratio $\chi^{1/4} / \sigma^{1/2}$ is dominated by $\sigma$ (c.f.\ Fig.~\ref{FIG_014}b). The range of values for $\lambda \geq 1.00$, i.e.\ for significantly overlapping pseudoparticles, $\chi^{1/4} / \sigma^{1/2} = 0.41 \ldots 0.28$, is of the right order of magnitude when compared to the lattice result \\ $(\chi^{1/4} / \sigma^{1/2})_\textrm{lattice} = 0.486 \pm 0.010$ \cite{Teper:1998kw}.

We have also calculated quark antiquark potentials for different values of $\lambda$ (c.f.\ section~\ref{SEC_005}). The results are shown as functions of the quark antiquark separation in Fig.~\ref{FIG_014}c. For $\lambda \geq 1.25$ the potential is clearly confining, whereas for $\lambda \leq 0.50$ it is unambiguously not confining, a result which is in agreement with previous results from this section.


\subsubsection{``Singular gauge pseudoparticles'' of different size}

\begin{figure}[b]
\begin{center}
\includegraphics{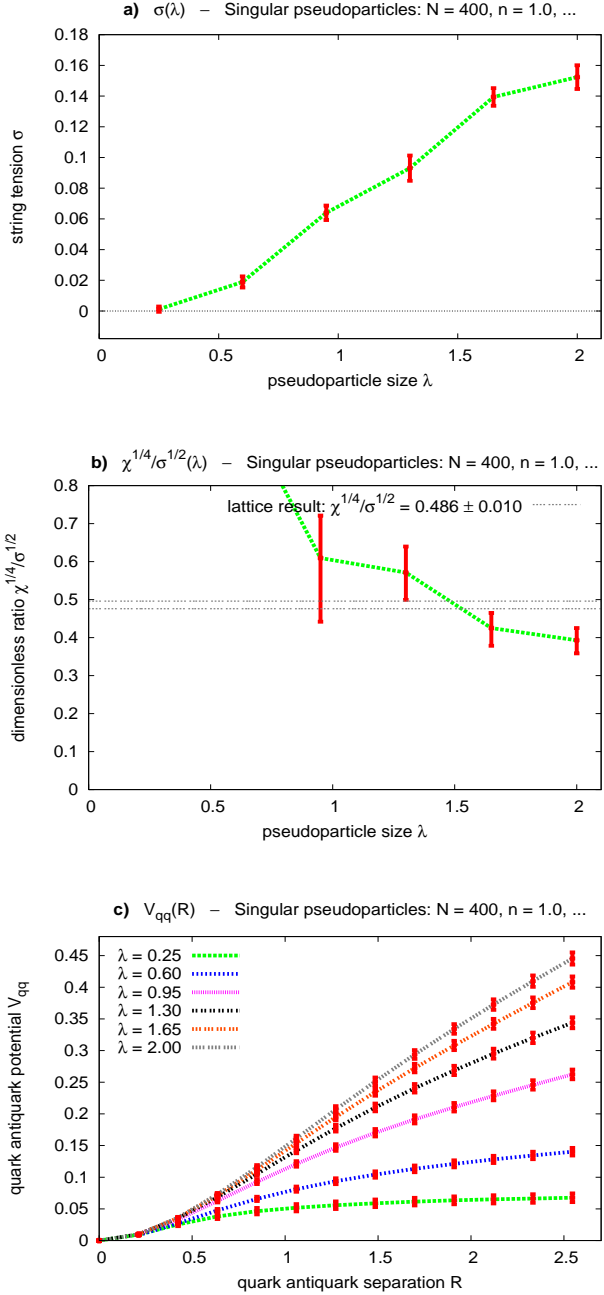}
\caption{\label{FIG_015}Singular gauge pseudoparticles, $N = 400$, $n = 1.0$, $\epsilon = 0.1$, $g = 4.0$.
\textbf{a)}~$\sigma$ plotted against $\lambda$.
\textbf{b)}~$\chi^{1/4} / \sigma^{1/2}$ plotted against $\lambda$.
\textbf{c)}~$V_{\textrm{q} \bar{\textrm{q}}}$ for different pseudoparticle sizes plotted against the separation of the quarks.
}
\end{center}
\end{figure}
We have also studied ensembles of singular gauge instantons, antiinstantons and akyrons. These ``singular gauge pseudoparticles'' are obtained by replacing $1 / (x^2 + \lambda^2)$ in (\ref{EQN_001}) to (\ref{EQN_003}) by $\lambda^2 / (x^2 + \epsilon^2) (x^2 + \lambda^2)$, where $\epsilon$ is an ultraviolet regulator, which has been included due to numerical reasons.

We consider ensembles with $N = 400$, $n = 1.0$, $\epsilon = 0.1$, $g = 4.0$ and $\lambda \in \{0.25 \, , \, 0.60 \, , \, \ldots \, , \, 2.00\}$.

Numerical results (c.f.\ Fig.~\ref{FIG_015}) and their interpretation are similar to those obtained with Gaussian localized pseudoparticles. Of course, this is what one would expect, because the gauge fields of singular gauge pseudoparticles decrease much faster than those of regular gauge pseudoparticles, i.e.\ like $1/|x|^3$ instead of $1/|x|$. Therefore, singular gauge pseudoparticles are quite similar to Gaussian localized pseudoparticles with respect to long range interactions.

The result that there is no confinement for ensembles of singular gauge pseudoparticles of small size ($\lambda/\bar{d} \ltapprox 0.6$) is in agreement with what has been obtained in instanton gas and instanton liquid models, where typically $\lambda/\bar{d} \approx 1/3$ (c.f.\ e.g.\ \cite{Schafer:1996wv}). Furthermore, it has been proposed that large size singular instantons can produce confinement (\cite{Diakonov:1996} quoted in \cite{Schafer:1996wv}), which is also supported by our findings.


\subsection{\label{SEC_010}Instantons and antiinstantons versus akyrons}

In the following we present evidence that confinement arises due to instantons and antiinstantons, whereas akyrons do not produce confinement. To this end, we consider ensembles with the same number of pseudoparticles but different ratios $N_\textrm{instanton} : N_\textrm{antiinstanton} : N_\textrm{akyron}$. In detail, we compare the following ensembles with \\ $N = 400$, $n = 1.0$, $\lambda = 0.5$ and $g = 4.0$:
\begin{itemize}
\item ``Akyron ensemble'': an ensemble containing $400$ akyrons (no instantons or antiinstantons).

\item ``Standard ensemble'': an ensemble containing $150$ instantons, $150$ antiinstantons and $100$ akyrons.

\item ``Instanton ensemble'': an ensemble containing $200$ instantons and $200$ antiinstantons (no akyrons).
\end{itemize}

\begin{figure}[t]
\begin{center}
\includegraphics{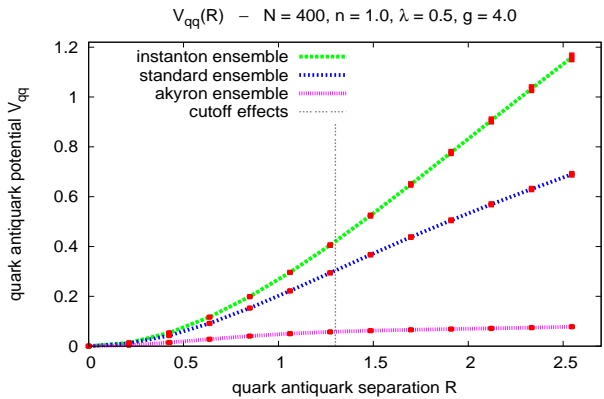}
\caption{\label{FIG_016}akyron ensemble ($400$ akyrons), standard ensemble ($150$ instantons, $150$ antiinstantons and $100$ akyrons) and instanton ensemble ($200$ instantons and $200$ antiinstantons), $N = 400$, $n = 1.0$, $\lambda = 0.5$, $g = 4.0$. $V_{\textrm{q} \bar{\textrm{q}}}$ plotted against the separation of the quarks.}
\end{center}
\end{figure}

Fig.~\ref{FIG_016} shows the quark antiquark potential as a function of the separation for all three ensembles (c.f.\ section~\ref{SEC_005}).  The standard ensemble and the instanton ensemble exhibit a confining potential, whereas the akyron curve indicates that there is no confinement.

We have obtained numerical values for the string tension $\sigma$ via generalized Creutz ratios (c.f.\ section~\ref{SEC_004}). The results are $\sigma_\textrm{akyron} = 0.019 \pm 0.08$, \\ $\sigma_\textrm{standard} = 0.236 \pm 0.013$ and $\sigma_\textrm{instanton} = 0.512 \pm 0.023$. \\ The ratio of these values is given by \\ $\sigma_\textrm{akyron} : \sigma_\textrm{standard} : \sigma_\textrm{instanton} \approx 1 : 12 : 27$.

To compare dimensionless ratios, we have calculated the topological susceptibility $\chi$ and the critical temperature $T_\textrm{critical}$ in the standard ensemble and in the instanton ensemble: \vspace{0.3cm} \\
$(\chi^{1/4} / \sigma^{1/2})_\textrm{standard} = 0.35$, \\
$(\chi^{1/4} / \sigma^{1/2})_\textrm{instanton} = 0.26$, \vspace{0.2cm} \\
$(T_\textrm{critical} / \sigma^{1/2})_\textrm{standard} = 0.61$, \\
$(T_\textrm{critical} / \sigma^{1/2})_\textrm{instanton} = 0.59$. \vspace{0.3cm} \\
Obviously, akyrons increase the topological susceptibility, while they do not affect the critical temperature. Since $(\chi^{1/4} / \sigma^{1/2})_\textrm{standard} = 0.35$ is closer to the lattice result $(\chi^{1/4}/\sigma^{1/2})_\textrm{lattice} = 0.486 \pm 0.010$ \cite{Teper:1998kw} than \\ $(\chi^{1/4} / \sigma^{1/2})_\textrm{instanton} = 0.26$, it is beneficial with respect to quantitative results to consider ensembles containing not only instantons and antiinstantons but also akyrons.

The dimensionless ratios $\chi^{1/4} / \sigma^{1/2}$ and $T_\textrm{critical} / \sigma^{1/2}$ in the akyron ensemble are not meaningful: the topological susceptibility $\chi$ vanishes identically (c.f.\ Appendix~\ref{SEC_013}) and the ensemble average of the Polyakov loop $\langle L \rangle_\beta$ is close to $1$ even for very large values of $\beta$, i.e.\ there is no sign of a confinement deconfinement phase transition. From that and the fact that the string tension is more than ten times smaller in the akyron ensemble than in the other two ensembles we conclude that gauge field configurations made up solely of akyrons are not suited to produce confinement. On the other hand, gauge field configurations, which are responsible for confinement, necessarily contain instantons and antiinstantons.

It is interesting to note that any linear superposition of akyrons and, therefore, any field configuration in a pure akyron ensemble have zero topological charge density (c.f.\ Appendix~\ref{SEC_013}). Since akyron ensembles do not exhibit confinement, this supports the common expectation that confinement and topological charge are closely related.


\section{\label{SEC_011}Summary and outlook}

In this work we have presented the pseudoparticle approach, a numerical method to compute path integrals in effective SU(2) Yang-Mills theories. We have calculated the static quark antiquark potential at zero and at finite temperature, the topological susceptibility and the critical temperature of the confinement deconfinement phase transition in different pseudoparticle ensembles. The pseudoparticle approach is able to reproduce many essential features of SU(2) Yang-Mills theory with a comparatively small number of degrees of freedom.


\subsection{The pseudoparticle approach as a successful effective theory}

When using $400$ instantons, antiinstantons and akyrons, the static quark antiquark potential is linear for large separations with an attractive $1/R$-correction, which is in qualitative agreement with the bosonic string picture and with results from lattice calculations. The string tension $\sigma$ is not only positive but also an increasing function of the coupling constant $g$. Therefore, when the scale is set by the string tension, one can adjust the size of the spacetime hypersphere in physical units by choosing an appropriate value for $g$.

We have also calculated the the topological susceptibility $\chi$ and the critical temperature $T_\textrm{critical}$. The dimensionless ratios $\chi^{1/4} / \sigma^{1/2}$ and $T_\textrm{critical} / \sigma^{1/2}$ are constant for a wide range of coupling constants, i.e.\ $\sigma$, $\chi$ and $T_\textrm{critical}$ exhibit consistent scaling behaviors with respect to $g$. This success strongly indicates that the pseudoparticle approach has the potential to reproduce correct Yang-Mills physics. The values of both dimensionless ratios are of the right order of magnitude compared to lattice results.


\subsection{Properties of confining gauge field configurations}

For ensembles made up of instantons, antiinstantons and akyrons the string tension shows only a weak dependence on the pseudoparticle size $\lambda$. It seems that confinement is mainly a consequence of the long range behavior of the building blocks, which is unaffected by $\lambda$. This has been confirmed by considering ensembles of Gaussian localized pseudoparticles and ensembles of singular gauge pseudoparticles, for which the size parameter $\lambda$ strongly affects the long range behavior. For small $\lambda$ there is only little overlap between neighboring pseudoparticles and there is no sign of confinement. Increasing $\lambda$ to a value, where pseudoparticles overlap and interact significantly, restores quark confinement. The conclusion is that gauge field configurations, which are responsible for confinement, contain extended structures. On the other hand, gauge field configurations with only localized excitations do not produce confinement.

Comparing our ``standard ensemble'' with a pure instanton and antiinstanton ensemble and a pure akyron ensemble has shown that confinement arises due to instantons and antiinstantons and not because of akyrons. Keeping in mind that gauge field configurations made up solely of akyrons have vanishing topological charge density, our findings support the common expectation that topological charge and confinement are closely related. For quantitative results akyrons seem to play an important role. The dimensionless ratio $\chi^{1/4}/\sigma^{1/2}$ is significantly closer to the lattice result, when there are not only instantons and antiinstantons but also akyrons.


\subsection{Outlook}

An obvious possibility for future research is to calculate correlation functions in order to extract glueball masses. Comparing the resulting masses with results from lattice calculations would be another check of the pseudoparticle approach. Furthermore we plan to compute other observables at finite temperature, e.g.\ the energy density and the pressure.

A major new direction is to include fermions in the pseudoparticle approach. The goal would be to obtain a model for SU(2) Yang-Mills theory, which exhibits both chiral symmetry breaking and a confinement deconfinement phase transition.


\begin{acknowledgments}

It is a pleasure to thank Frieder Lenz for lots of inspiring discussions as well as for making many important and helpful suggestions. Furthermore, I would like to acknowledge useful conversations with Philippe~de~Forcrand, John~W.~Negele and Jan~M.~Pawlowski.

\end{acknowledgments}


\appendix

\section{\label{SEC_012}Color orientation matrices}

A color orientation is a spacetime independent gauge transformation. It can be specified by an element of SU(2): $U = c_0 + i c_a \sigma^a$ with $c_0^2 + \mathbf{c}^2 = 1$. Applying such a gauge transformation to a gauge field $A_\mu$ yields
\begin{eqnarray}
A_{\mu}' \ \ = \ \ U A_{\mu} U^{-1}
\end{eqnarray}
or expressed in components
\begin{eqnarray}
\nonumber & & \hspace{-0.44cm} A_{\mu}^a{}' \ \ = \ \ \textrm{Tr}(\sigma^a A_{\mu}') \ \ = \\
 & & = \ \ \underbrace{\left(\delta^{a b} \Big(c_0^2 - \mathbf{c}^2\Big) + 2 c_a c_b + \epsilon^{a b c} 2 c_0 c_c\right)}_{= \mathcal{C}^{a b}} A_{\mu}^b .
\end{eqnarray}
We refer to $\mathcal{C}^{a b}$ as color orientation matrix. Color orientation matrices fulfill
\begin{eqnarray}
 & & \hspace{-0.44cm} \mathcal{C} \mathcal{C}^T \ \ = \ \ \mathcal{C}^T \mathcal{C} \ \ = \ \ 1 \\
 & & \hspace{-0.44cm} \det(\mathcal{C}) \ \ = \ \ 1 .
\end{eqnarray}
Therefore, they are elements of SO(3).


\section{\label{SEC_013}Any linear superposition of akyrons has vanishing topological charge density}

The gauge field of a single akyron with index $i$, amplitude $\mathcal{A}(i)$, color orientation matrix $\mathcal{C}^{a b}(i)$ and position $z(i)$ is given by
\begin{eqnarray}
\nonumber & & \hspace{-0.44cm} A_\mu^a(i) \ \ = \ \ \mathcal{A}(i) \mathcal{C}^{a b}(i) a_{\textrm{akyron},\mu}^b(x-z(i)) \ \ = \\
 & & = \ \ \mathcal{A}(i) \mathcal{C}^{a 1}(i) \frac{x_\mu-z_\mu(i)}{(x-z(i))^2 + \lambda^2}
\end{eqnarray}
(c.f.\ (\ref{EQN_003})). In the corresponding field strength the derivative terms cancel each other:
\begin{eqnarray}
 & & \hspace{-0.44cm} F_{\mu \nu}^a(i) \ \ = \ \ \underbrace{\partial_\mu A_\nu^a(i) - \partial_\nu A_\mu^a(i)}_{= 0} + \epsilon^{a b c} A_\mu^b(i) A_\nu^c(i) .
\end{eqnarray}
For any linear superposition of akyrons
\begin{eqnarray}
A_\mu^a \ \ = \ \ \sum_i A_\mu^a(i)
\end{eqnarray}
the same is true:
\begin{eqnarray}
F_{\mu \nu}^a \ \ = \ \ \underbrace{\partial_\mu A_\nu^a - \partial_\nu A_\mu^a}_{= 0} + \epsilon^{a b c} A_\mu^b A_\nu^c .
\end{eqnarray}
For the topological charge density follows
\begin{eqnarray}
\nonumber & & \hspace{-0.44cm} q \ \ = \ \ \frac{1}{32 \pi^2} F_{\mu \nu}^a \tilde{F}_{\mu \nu}^a \ \ = \ \ \frac{1}{64 \pi^2} \epsilon_{\mu \nu \alpha \beta} F_{\mu \nu}^a F_{\alpha \beta}^a \ \ = \\
\nonumber & & = \ \ \frac{1}{64 \pi^2} \epsilon_{\mu \nu \alpha \beta} \epsilon^{a b c} \epsilon^{a d e} A_\mu^b A_\nu^c A_\alpha^d A_\beta^e \ \ = \\
 & & = \ \ \frac{1}{32 \pi^2} \underbrace{\epsilon_{\mu \nu \alpha \beta} (A_\mu^b A_\alpha^b)}_{= 0} (A_\nu^c A_\beta^c) \ \ = \ \ 0 .
\end{eqnarray}


\section{\label{SEC_13}Instantons and antiinstantons form transverse gauge fields, akyrons form longitudinal gauge fields}

Any gauge field $A_\mu^a$ can be written as a sum of plane waves:
\begin{eqnarray}
A_\mu^a(x) \ \ = \ \ \frac{1}{(2 \pi)^4} \int d^4k \, e^{-i k x} \tilde{A}_\mu^a(k) ,
\end{eqnarray}
where $\tilde{A}_\mu^a$, the Fourier transform of $A_\mu^a$, is given by
\begin{eqnarray}
\tilde{A}_\mu^a(k) \ \ = \ \ \int d^4k \, e^{i k x} A_\mu^a(x) .
\end{eqnarray}

The Fourier transformed gauge field $\tilde{A}_\mu^a$ can be decomposed in a transverse and a longitudinal part:
\begin{eqnarray}
\tilde{A}_\mu^a(k) \ \ = \ \ \tilde{A}_{\mu,\textrm{transverse}}^a(k) + \tilde{A}_{\mu,\textrm{longitudinal}}^a(k)
\end{eqnarray}
with
\begin{eqnarray}
\label{EQN_019} & & \hspace{-0.44cm} k_\mu \tilde{A}_{\mu,\textrm{transverse}}^a(k) \ \ = \ \ 0 \\
\label{EQN_020} & & \hspace{-0.44cm} \tilde{A}_{\mu,\textrm{longitudinal}}^a(k) \ \ \propto \ \ k_\mu .
\end{eqnarray}

Superpositions of instantons (\ref{EQN_001}) and antiinstantons (\ref{EQN_002}) form transverse gauge fields, whereas superpositions of akyrons (\ref{EQN_003}) form longitudinal gauge fields. This can be seen by considering the Fourier transforms of these pseudoparticles:
\begin{eqnarray}
\label{EQN_021} & & \hspace{-0.44cm} \tilde{a}_{\mu,\textrm{instanton}}^a(k) \ \ = \ \ \eta_{\mu \nu}^a k_\nu f(|k|) \\
\label{EQN_022} & & \hspace{-0.44cm} \tilde{a}_{\mu,\textrm{antiinstanton}}^a(k) \ \ = \ \ \bar{\eta}_{\mu \nu}^a k_\nu f(|k|) \\
\label{EQN_023} & & \hspace{-0.44cm} \tilde{a}_{\mu,\textrm{akyron}}^a(k) \ \ = \ \ \delta^{a 1} k_\mu f(|k|) ,
\end{eqnarray}
where
\begin{eqnarray}
\label{EQN_024} f(k) \ \ = \ \ \frac{8 \pi^2 i}{|k|^4} \left(\frac{|k| \lambda K_1(|k| \lambda)}{2} - \frac{k^2 \lambda^2 K_1'(|k| \lambda)}{2}\right)
\end{eqnarray}
($K_1$ is a modified Bessel function of imaginary argument). (\ref{EQN_021}) and (\ref{EQN_022}) satisfy (\ref{EQN_019}) due to the antisymmetry of $\eta_{\mu \nu}^a$ and $\bar{\eta}_{\mu \nu}^a$, while (\ref{EQN_023}) obviously fulfills (\ref{EQN_020}).


\section{\label{SEC_014}Instantons and akyrons form a basis of all gauge field configurations}

In this appendix we show that ``almost any gauge field configuration'' can be represented by a linear superposition of infinitely many instantons and akyrons.

The starting point is the ``continuum limit'' of (\ref{EQN_005}) without antiinstantons:
\begin{eqnarray}
\nonumber & & \hspace{-0.44cm} A_\mu^a(x) \ \ = \\
\nonumber & & = \ \ \int d^4z \, \Bigg(\sum_{i=1}^9 \mathcal{A}(i,z) \mathcal{C}^{a b}(i,z) a_{\mu,\textrm{instanton}}^b(x-z) + \\
\label{EQN_025} & & \hspace{0.62cm} \sum_{j=10}^{12} \mathcal{A}(j,z) \mathcal{C}^{a b}(j,z) a_{\mu,\textrm{akyron}}^b(x-z)\Bigg)
\end{eqnarray}
(the sum over all pseudoparticles has been replaced by an integration over spacetime; furthermore, nine instantons and three akyrons are allowed to share the same position).

Inserting (\ref{EQN_001}) and (\ref{EQN_003}) in (\ref{EQN_025}) yields
\begin{eqnarray}
\nonumber & & \hspace{-0.44cm} A_\mu^a(x) \ \ = \ \ \int d^4z \, \Bigg(\underbrace{\sum_{i=1}^9 \mathcal{A}(i,z) \mathcal{C}^{a b}(i,z)}_{= \mathcal{S}^{a b}(z)} \eta_{\mu \nu}^b + \\
\label{EQN_026} & & \hspace{0.62cm} \underbrace{\sum_{j=10}^{12} \mathcal{A}(j,z) \mathcal{C}^{a b}(j,z) \delta^{b 1}}_{= \mathcal{S}^{a 0}(z)} \delta_{\mu \nu}\Bigg) \frac{x_\nu - z_\nu}{(x-z)^2 + \lambda^2} .
\end{eqnarray}
It can be shown that in general nine color orientation matrices form a basis of all $3 \times 3$-matrices \cite{Wagner:2006}. Therefore, any $\mathcal{S}^{a b}$ and $\mathcal{S}^{a 0}$ can be realized by suitably chosen amplitudes $\mathcal{A}(i,z)$ and $\mathcal{A}(j,z)$. Hence, the problem has been reduced to the question whether any gauge filed configuration $A_\mu^a$ can be represented by suitably chosen $\mathcal{S}^{a b}$ and $\mathcal{S}^{a 0}$.

Fourier transforming (\ref{EQN_026}) turns the convolution into an ordinary multiplication:
\begin{eqnarray}
\nonumber & & \hspace{-0.44cm} \tilde{A}_\mu^a(k) \ \ = \\
\nonumber & & = \ \ \Big(\tilde{\mathcal{S}}^{a b}(k) \eta_{\mu \nu}^b + \tilde{\mathcal{S}}^{a 0}(k) \delta_{\mu \nu}\Big) \int d^4x \, e^{i k x} \frac{x_\nu}{x^2 + \lambda^2} \ \ = \\
 & & = \ \ \Big(\tilde{\mathcal{S}}^{a b}(k) \eta_{\mu \nu}^b + \tilde{\mathcal{S}}^{a 0}(k) \delta_{\mu \nu}\Big) k_\nu f(k)
\end{eqnarray}
($f$ is defined by (\ref{EQN_024})).

Without loss of generality we consider $a = 1$:
\begin{eqnarray}
\nonumber & & \hspace{-0.44cm}
\left(\begin{array}{c} \tilde{A}_0^1(k) \\ \tilde{A}_1^1(k) \\ \tilde{A}_2^1(k) \\ \tilde{A}_3^1(k) \end{array}\right)
\ \ = \\
\label{EQN_027} & & = \ \ f(k)
\underbrace{\left(\begin{array}{cccc}
k_0 & -k_1 & -k_2 & -k_3 \\
k_1 &  k_0 & -k_3 &  k_2 \\
k_2 &  k_3 &  k_0 & -k_1 \\
k_3 & -k_2 &  k_1 &  k_0 \\
\end{array}\right)}_{= \mathcal{K}(k)}
\left(\begin{array}{c} \tilde{\mathcal{S}}^{1 0}(k) \\ \tilde{\mathcal{S}}^{1 1}(k) \\ \tilde{\mathcal{S}}^{1 2}(k) \\ \tilde{\mathcal{S}}^{1 3}(k) \end{array}\right)
.
\end{eqnarray}
For $k \neq 0$ this equation can be solved for \\ $(\tilde{\mathcal{S}}^{1 0},\tilde{\mathcal{S}}^{1 1},\tilde{\mathcal{S}}^{1 2},\tilde{\mathcal{S}}^{1 3})$, because $f \neq 0$ and \\ $\det(\mathcal{K}) = |k|^4 \neq 0$.  For $k = 0$ both $f$ and $\mathcal{K}$ are singular. To study this case, we first deduce
\begin{eqnarray}
\label{EQN_028} \int d^4x \, A_\mu^a(x) \ \ = \ \ 0
\end{eqnarray}
from (\ref{EQN_026}) by applying a proper regularization scheme. (\ref{EQN_028}) implies $\tilde{A}_\mu^a(k=0) = 0$. Inserting this in (\ref{EQN_027}) \\ shows that the value of $\tilde{\mathcal{S}}^{a B}(k = 0)$ has no effect on the gauge field $A_\mu^a$. On the other hand, changing $\tilde{\mathcal{S}}^{a B}(k = 0)$ amounts to a constant shift of $\mathcal{S}^{a B}$: $\mathcal{S}^{a B} \rightarrow \mathcal{S}^{a B} + \mathcal{S}_0^{a B}$. That is adding $\mathcal{S}_0^{a B}$ changes $\mathcal{S}^{a B}$, whereas the gauge field $A_\mu^a$ remains unaltered. To get rid of this redundancy, we require
\begin{eqnarray}
\label{EQN_029} \int d^4x \, \mathcal{S}^{a B}(x) \ \ = \ \ 0 .
\end{eqnarray}
To be able to represent any gauge field configuration $A_\mu^a$, we have to find a way around (\ref{EQN_028}). This can easily be achieved by adding constants $B_\mu^a$ to (\ref{EQN_025}).

The final result is the following: any gauge field configuration $A_\mu^a$ has a unique expansion
\begin{eqnarray}
\nonumber & & \hspace{-0.44cm} A_\mu^a(x) \ \ = \\
\nonumber & & = \ \ \int d^4z \, \Bigg(\sum_{i=1}^9 \mathcal{A}(i,z) \mathcal{C}^{a b}(i,z) a_{\mu,\textrm{instanton}}^b(x-z) + \\
 & & \hspace{0.62cm} \sum_{j=10}^{12} \mathcal{A}(j,z) \mathcal{C}^{a b}(j,z) a_{\mu,\textrm{akyron}}^b(x-z)\Bigg) + B_\mu^a
\end{eqnarray}
in terms of $\mathcal{A}(i,z)$ and $\mathcal{A}(j,z)$ constrained by (\ref{EQN_029}) and $B_\mu^a$, where $\mathcal{C}^{a b}(i,z)$, $i=1,\ldots,9$, as well as $\mathcal{C}^{a 1}(j,z)$, \\ $j=10,\ldots,12$ are linearly independent color orientation matrices.



\end{document}